\documentclass[fleqn,usenatbib]{mnras}

\usepackage{newtxtext,newtxmath}
\usepackage{pdfcomment}

\usepackage[T1]{fontenc}

\DeclareRobustCommand{\VAN}[3]{#2}
\let\VANthebibliography\thebibliography
\def\thebibliography{\DeclareRobustCommand{\VAN}[3]{##3}\VANthebibliography}

\usepackage{graphicx}	
\usepackage{amsmath}	

\providecommand{\aap}{Astronomy \& Astrophysics}

\providecommand{\apjl}{The Astrophysical Journal Letters}

\newcommand{\resp}[1]{\textcolor{black}{#1}}
\usepackage{graphicx}
\usepackage{subcaption} 

\title[Dust and radicals of E1]{Dust and coma radicals in C/2024 E1 (Wierzchos): \resp{ evolution of the activity}}

\author[I. Mariblanca-Escalona et al]{
Irene Mariblanca-Escalona,$^{1}$\thanks{E-mail: irenem@iaa.csic.es}
Pedro J. Gutiérrez$^{1}$
Fernando Moreno,$^{1}$
Luisa M. Lara$^{1}$
\\
$^{1}$Instituto de Astrofísica de Andalucía, CSIC, Glorieta de la Astronomía s/n, E-18008 Granada, Spain
}

\date{Accepted 2026 June 11. Received 2026 June 11; in original form 2026 May 13}

\pubyear{\the\year{}}

\begin{document}
\label{firstpage}
\pagerange{\pageref{firstpage}--\pageref{lastpage}}
\maketitle

\begin{abstract} 
We present multi-epoch R-band imaging and long-slit spectroscopy of the dynamically new comet C/2024 E1 (Wierzchos) during its pre-perihelion phase, spanning heliocentric distances from 4.5 to 2.3 au. The most prominent fluorescence emissions in the visible wavelength range are detected. We measure production rates of CN, C$_2$, and C$_3$, while no significant NH$_2$ signal is identified. 
\resp{In our dataset, CN became detectable at 3.48~au, and C$_3$ and C$_2$ at 2.80~au.} The mean ratios $\log\mathrm{C}_2/\mathrm{CN} = 0.00 \pm 0.07$ and $\log\mathrm{C}_3/\mathrm{CN} = -1.03 \pm 0.09$ place the comet in an intermediate taxonomic regime between typical and carbon-depleted populations. Dust production rates, ejection velocities, and particle size distributions are constrained using a Monte Carlo dust tail model that reproduces the observed coma morphology. The dust population is characterized by a power-law size distribution with an exponent of $\sim -3.5$, and particle sizes ranging from 1~$\mu$m to 4~mm. The inferred initial ejection velocities span from $\sim$20~m~s$^{-1}$ to $\sim$100~m~s$^{-1}$. We use the empirical relation between CN and OH to estimate the water production rate and derive the dust-to-gas mass ratio for heliocentric distances $r_{\rm{h}} \lesssim 3$~au, where water sublimation is expected to dominate the activity regime. The dust-to-gas mass ratio remains consistently above unity, ranging from $\sim$1 to 3 depending on the choice of scale lengths and the CN--OH conversion factor. This result characterizes C/2024 E1 as a ``dust-rich'' comet, where the mass loss is dominated by the refractory component rather than volatiles.
\end{abstract}

\begin{keywords}
methods: numerical -- comets: general -- comets: individual: C/2024 E1
 
\end{keywords}


\section{Introduction}
Comets are remnants of the Solar System's formation, composed of refractory material and volatile ices \citep{1950ApJ...111..375W}. While water ice sublimation typically governs activity in the inner Solar System, species with higher volatility, such as CO and CO$_2$, are responsible for driving it at larger heliocentric distances \citep[see e.g.][] {2017ApJ...849L...8M}. Given the difficulty in observing these primary volatiles from the ground, cometary activity in the visible range is frequently characterized by the study of secondary radicals (e.g., CN, C$_2$, C$_3$ and NH$_2$) arising from the photodissociation of more complex parent molecules. 

Among the different cometary families \citep{1996ASPC..107..173L}, Dynamically New Comets (DNCs) are particularly valuable. They make their first passage through the inner Solar System after long-term storage in the Oort cloud \citep{2004ASPC..323..371D}. Thus, it is thought they possess surfaces with limited thermal processing and thus probably preserve more pristine volatile reservoirs than evolved comets.

Large observational surveys in the visible range have established the general properties of cometary activity by analysing samples of tens to hundreds of objects \citep[e.g.,][]{1995Icar..118..223A, 2009Icar..201..311F, 2011Icar..213..280L, 2012Icar..218..144C, 2025PSJ.....6..248B}, providing a comparative framework for gas composition and activity levels. However, these surveys are strongly biased toward heliocentric distances $r_{\rm{h}} \lesssim 3$~au, where activity is more easily detectable. This bias is especially critical for DNCs, as the lack of observations at larger distances limits our understanding of the onset of activity of their volatiles. While the behaviour of comets in this distant regime still remains poorly constrained, an increasing number of objects have recently been observed to exhibit significant activity well beyond 3~au. This includes not only extremely active comets such as the famous Hale–Bop, which showed visible emission from the aforementioned radicals before 4~au \citep{2003A&A...397.1109R}, but also more moderate objects like C/2017 K2 \citep{2025PSJ.....6..272C} and C/2020 V2 \citep{2025MNRAS.543.1178A}, where these emissions has also been detected at distances greater than 3~au. These studies underscore the necessity of characterizing cometary activity in this poorly explored regime. 

Such an observational requirement is further highlighted by the upcoming ESA/JAXA \textit{Comet Interceptor} mission \citep{2024SSRv..220....9J}. Since the mission aims to encounter a DNC at $r_{\rm{h}} \sim 1$~au, its success relies on identifying and monitoring targets while they are still far from the Sun. Consequently, both the optimization of encounter strategies and the eventual interpretation of \textit{in situ} data depend on a robust ground-based understanding of analogous objects over a wide range of heliocentric distances. Within this framework, extending the observational coverage of DNCs beyond $r_{\rm{h}} \sim 3$~au is essential to constrain the onset of activity and provide the empirical context required to interpret the diverse activity regimes observed in these primitive bodies.

C/2024 E1 (Wierzchos; hereafter E1) represents a favourable target to investigate the evolution of cometary activity over a wide range of heliocentric distances. The comet was discovered by the Mt. Lemmon Survey on 2024 March 3, with prediscovery observations from the Zwicky Transient Facility (ZTF) extending the observational arc back to 2024 February 15. E1 follows a long-period orbit consistent with an Oort cloud origin ($e \sim 1.0$, $q \sim 0.56$~au, $i \sim 75.2^{\circ}$) \citep[][]{2024CBET.5364....1W}. At the time of its earliest detection, the object was at $r_{\rm{h}} \sim 8.2$~au enabling its activity to be monitored over an extended inbound trajectory of nearly two years prior to perihelion. At that distance, it already exhibited a condensed 4'' coma, providing a valuable opportunity to track the development of activity from the distant regime into the inner Solar System.

Shortly after discovery, E1 was observed with the \textit{James Webb Space Telescope} (JWST) in the near-infrared when it was at $r_{\rm{h}} \sim 7$~au \citep{2025MNRAS.541L...8S}. These authors found clear absorption features attributed to water ice grains and CO$_2$ emission, which appears to drive distant activity. Surprisingly, no CO emission or absorption features were detected. This is particularly notable for a comet observed at such a large heliocentric distance, as CO is expected to be a dominant volatile species driving activity in the distant regime. In several Oort Cloud comets observed at similar or closer distances, such as C/1995 O1 (Hale–Bopp) at 6.6~au and C/2006 W3 (Christensen) at $\sim$5~au, CO outgassing has been identified as a primary engine of activity \citep{1997Sci...275.1915B, 2010A&A...518L.149B}. Even when CO$_2$ begins to dominate the outgassing between 6 and 8~au, the lack of CO detections in E1 remains unusual, as this species is typically present in this heliocentric range \citep[e.g.,][]{2022DPS....5441203R}. 

The comet reached perihelion on 2026 January 26, and subsequent observations indicate a rapid decrease in brightness and the disappearance of a centrally condensed source, consistent with the disruption of the nucleus shortly after perihelion \citep[][]{2026CBET.5669....1Y}. In this work, we present \resp{a} 
ground-based spectroscopic time series of E1 obtained prior to perihelion. The dataset consists of multi-epoch long-slit visible spectroscopy and R-band imaging acquired between 4.5 and 2.3~au using CAFOS at the Calar Alto Observatory. The monitoring was carried out until the comet became unobservable from the Northern hemisphere due to solar conjunction.

\section{OBSERVATIONS AND DATA REDUCTION}

E1 was observed by using CAFOS instrument mounted on the 2.2~m telescope at Calar Alto Observatory (Almería, Spain), between 2025 April 1 and 2025 September 19, covering the pre-perihelion phase from approximately 4.5 to 2.3~au heliocentric distances. Data were acquired at intervals of 0.1--0.2~au along the orbit, providing a good sampling of the activity evolution.  At each epoch, taking advantage of the dual imaging and spectroscopic capabilities of CAFOS, we obtained R-band images and long-slit visible spectra. CAFOS is equipped with a $2048 \times 2048$ pixel CCD detector providing a spatial scale of 0.53\arcsec\ pixel$^{-1}$. This corresponds to a physical resolution between 1705 and 884 km~pixel$^{-1}$, depending on the observing night\resp{, with an average seeing of 1.7\arcsec\ across our entire dataset}. Each observing run lasted between about 2~h and 1~h, decreasing as the comet approached the Sun. Table~\ref{tab:dataset} summarises the dataset.
 
\subsection{Imaging}

Imaging data were obtained using the R-band filter to trace the dust coma morphology. Multiple exposures were acquired each night (see Table~\ref{tab:dataset} for full details), ensuring adequate signal-to-noise while avoiding saturation. Although the CAFOS instrument has a nominal field of view of $18 \times 18\arcmin$, only the central $7\arcmin \times 7\arcmin$ region was used in our analysis, which was sufficient to fully encompass the comet’s coma and tail.

\begin{figure} 
    \centering
    \includegraphics[width=0.5\textwidth]{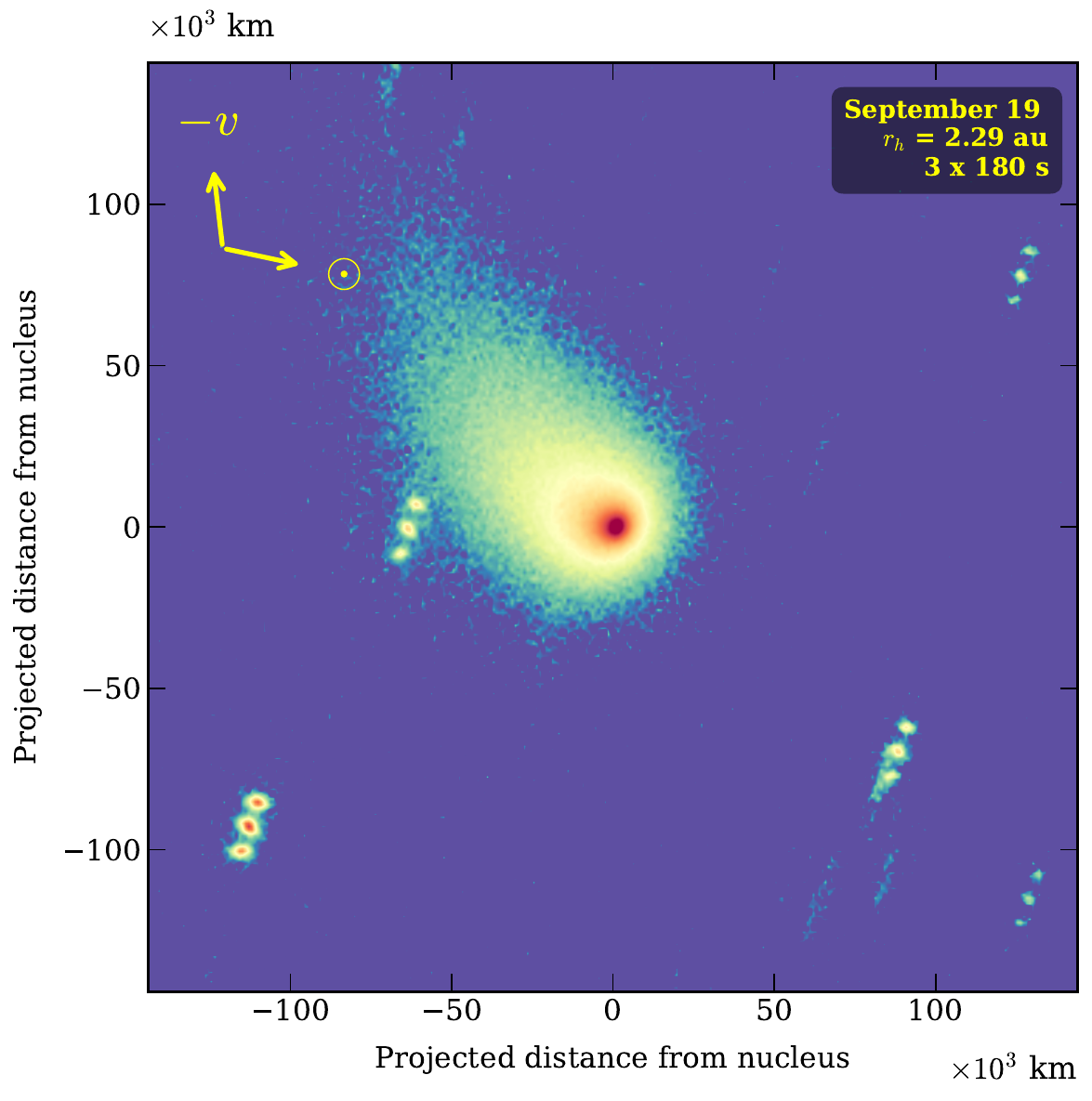} 
    \caption{R-band intensity map of comet E1 obtained on September 19, 2025 (see Table \ref{tab:dataset}). The $x$ and $y$ axes correspond to the projected distance from the nucleus in units of $10^{3}$\,km. Surface brightness is color-coded from 19 to 24\,mag/arcsec$^2$. North is up and East to the left. Arrows indicate two directions: the Sun vector pointing towards the Sun, and the anti-velocity vector, indicative of the dust tail direction.}
    \label{fig:image_r.pdf}
\end{figure}

Standard data reduction was performed in Python and included bias subtraction and flat-field correction. A World Coordinate System (WCS) solution was derived for each frame using \texttt{astrometry.net}. A background model was generated with \texttt{photutils} (\texttt{Background2D} combined with \texttt{MedianBackground}) after masking sources via sigma-clipping, resulting in a synthetic sky frame that was subtracted from the science images.  A nightly zero point was derived for each frame by comparison with catalogue magnitudes. Flux calibration was carried out by cross-matching field stars with the Pan-STARRS1 catalogue using \texttt{astroquery.XMatch} (within a $1\arcsec$ radius). Instrumental magnitudes were obtained through aperture photometry, and colour-term corrections were applied to convert Pan-STARRS $g-r$ colours to the Cousins $R$ band \citep{2012ApJ...750...99T}. Finally, individual frames were stacked using a median combination in the comet’s optocentre reference frame. The final stacked images were converted to units of mag~arcsec$^{-2}$. An example is shown in Figure~\ref{fig:image_r.pdf}.

\subsection{Long-slit spectroscopy}

\begin{figure*} 
    \centering
    

   \includegraphics[width=1\textwidth]{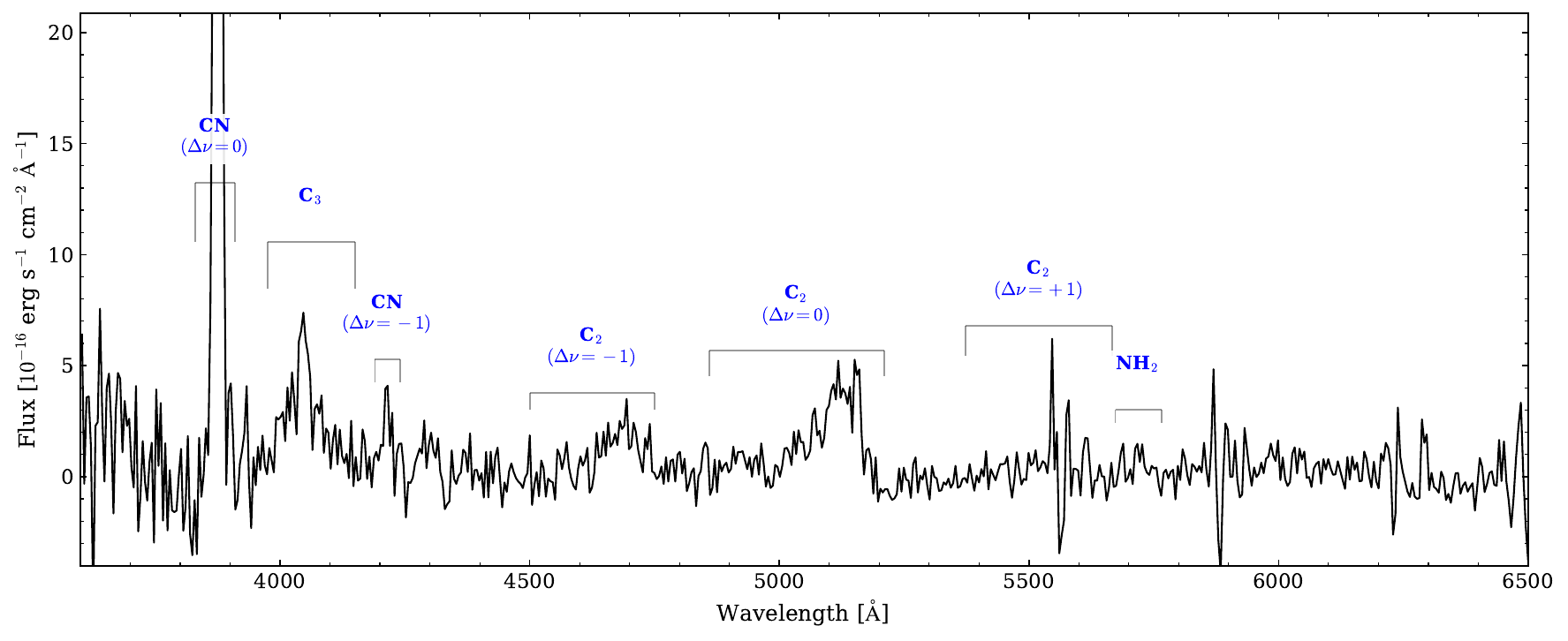} 
   \caption{1D spectrum of pure gas emission of E1 on 2025 September 19 with the CAFOS instrument (B200 grism). The spectra was obtained by subtracting the dust model from the observed 2D spectrum and integrating the residual emission within a spatial window of $\pm 50,000$~km from the nucleus. The most prominent cometary emission features in the visible range (CN, C$_2$, and C$_3$) are clearly detected.}
    \label{fig:1d_spectra.pdf}
\end{figure*}

Long-slit spectroscopic observations were obtained with the grism B400 for most of the campaign, covering approximately 3400--7400~\AA\ with a spectral dispersion close to $\sim$9.3~\AA~pixel$^{-1}$ in the regions of the main cometary emission bands (CN, C$_2$, C$_3$, and NH$_2$). For the final observation on 2025 September 19, we switched to the B200 grism, which provides higher spectral resolution with a dispersion of $\sim$4.7~\AA~pixel$^{-1}$ over 3500--8000~\AA, to explore the possible detection of the forbidden [O~I] emission line near 6300~\AA, a common proxy for H$_2$O dissociation. For all our observations, we used a 1.5\arcsec\ slit width for the comet and solar analogue stars to optimise spectral resolution while maintaining sufficient signal, whereas a wider 10\arcsec\ slit was used for spectrophotometric standard stars to minimise slit losses and ensure reliable flux calibration. \resp{The slit was oriented along the Sun-comet line as projected on the sky-plane.} Standard data reduction was performed in Python using available packages and customised routines. Master bias frames were subtracted from all images, and flat-field correction was applied using dome or twilight flats depending on the night. Cosmic-ray hits were removed with automated routines, and wavelength calibration was performed using HgHeRb arc-lamps, fitting a second-degree polynomial along the CCD to convert pixel positions to wavelengths. Flux calibration was performed using spectrophotometric standards BD+33\,2642 and P330E, observed at airmasses similar to that of the comet, ensuring similar extinctions. Since the comet did not fill the slit, clean sky regions at the edges of the CCD could be used to estimate and subtract the sky background. 
Finally, we isolated the gas emissions from the observed spectra, which contain contributions from both dust-scattered sunlight and gas. We used the following solar analogue stars: HD\,168009, TYC\,3544\,548\,1, and P330E, with the latter also serving as a spectrophotometric standard. This dual role of P330E allowed us to use it as both a solar analogue and a flux standard throughout all nights from July 13 onward, by adopting different slit widths: a 1.5'' slit for dust subtraction and a 10'' slit for flux calibration, aiming at an optimised and consistent observing strategy.

To model the dust continuum, we fitted third-degree polynomials to both the comet spectrum and the solar analogue stars in continuum regions free of gas emission (3750--3810, 3950--3975, 4760--4800, 5300--5350, 5850--5950 and 6450--6550\,\AA). The solar analogue spectrum was first normalized by its maximum to set its continuum to unity. We then scaled the normalized analogue by the ratio between the polynomial fits of the comet and the analogue, matching the comet’s dust continuum and accounting for reddening. Subtracting this scaled analogue from the observed spectrum removes the dust contribution, producing a flat, dust-subtracted spectrum suitable for the analysis of gas emission lines. An example of the resulting pure gas spectra is shown in Figure~\ref{fig:1d_spectra.pdf}. 

\begin{table*} 
\caption{Log of the observations. Columns list: mean UT of the spectra (date UT), exposure times for spectra and images, time to perihelion $\Delta t_{\rm{p}}$, heliocentric distance $r_{\rm{h}}$, geocentric distance $\Delta$, radial velocity $v$, position angle of the extended Sun-to-comet radius vector PsAng, position angle of the heliocentric velocity vector PsAMV, position angle of the projected nucleus-to-observer vector PlAng, and phase angle PhAng. For 2025-03-23 and 2025-08-27 the date UT corresponds to the mean of the image exposures. All geometric data were obtained for the UT date using the JPL Horizons Ephemeris Generator.}
\label{tab:dataset}
\centering
\begin{tabular}{c c c c c c c c c c c c}
\hline
Date UT & Spectra (s) & Images (s) & $\Delta t_{\rm{p}}$(days) & $r_{\rm{h}}$(au) & $\Delta$(au) & $v$(km s$^{-1}$) & PsAng($^{\rm{o}}$) & PsAMV($^{\rm{o}}$) & PlAng($^{\rm{o}}$) & PhAng($^{\rm{o}}$) \\
\hline
2025-03-23T04:21:46 & -------- & 4$\times$100 s & -303.59 & 4.47 & 4.42 & -18.62 & 272.21 & 318.40 & -12.21 & 12.85 \\
2025-04-01T03:37:51 & 3$\times$1400 s & 70 s + 3$\times$180 s & -294.62 & 4.38 & 4.28 & -18.79 & 264.31 & 316.08 & -13.05 & 13.20 \\
2025-04-29T03:34:58 & 2$\times$1400 s & 6$\times$100 s & -266.62 & 4.07 & 3.85 & -19.38 & 236.75 & 307.19 & -13.76 & 14.27 \\
2025-05-04T02:52:38 & 1200 s & 3$\times$180 s & -261.65 & 4.01 & 3.78 & -19.49 & 231.26 & 305.33 & -13.54 & 14.47 \\
2025-05-21T02:57:00 & 2$\times$1400 s & 180 s + 2$\times$300 s & -244.65 & 3.82 & 3.54 & -19.90 & 210.80 & 298.61 & -11.93 & 15.21 \\
2025-05-29T00:33:27 & 2$\times$1400 s & 3$\times$180 s & -236.74 & 3.73 & 3.44 & -20.10 & 200.44 & 295.65 & -10.73 & 15.60 \\
2025-06-19T01:22:49 & 2$\times$1500 s & 3$\times$180 s & -215.71 & 3.48 & 3.19 & -20.67 & 171.11 & 293.18 & -6.22  & 16.80 \\
2025-06-21T23:19:00  & 2$\times$1500 s  & 3$\times$180 s & -212.80 & 3.45 & 3.16 & -20.75 & 166.99 & 294.34 & -5.46  & 16.99 \\
2025-06-27T22:32:13 & 2$\times$1500 s & 3$\times$180 s & -206.83 & 3.37 & 3.09 & -20.92 & 158.63 & 298.99 & -3.81  & 17.40 \\
2025-07-01T23:32:01 & 2$\times$1200 s & 3$\times$180 s & -202.79 & 3.32 & 3.05 & -21.05 & 153.07 & 304.34 & -2.64  & 17.69 \\
2025-07-13T21:04:56 & 1500 s & 3$\times$180 s & -190.89 & 3.18 & 2.94 & -21.42 & 137.48 & 330.43 & 1.05 & 18.60 \\
2025-07-20T20:50:10 & 1500 s & -------- & -183.90 & 3.09 & 2.88 & -21.66 & 129.00 & 346.32 & 3.33 & 19.17 \\
2025-07-28T20:48:42 & 1500 s & 3$\times$180 s & -175.90 & 2.99 & 2.82 & -21.93 & 120.01 & 358.35 & 6.00 & 19.85 \\
2025-08-02T20:28:59 & 1500 s & -------- & -170.91 & 2.93 & 2.78 & -22.11 & 114.81 & 2.76 & 7.67 & 20.27 \\
2025-08-12T20:14:17 & 1500 s & 3$\times$180 s & -160.92 & 2.80 & 2.71 & -22.49 & 105.27 & 7.22 & 10.97 & 21.12 \\
2025-08-20T20:15:54 & 1500 s & 3$\times$180 s & -152.92 & 2.69 & 2.66 & -22.81 & 98.44 & 8.43 & 13.51 & 21.77 \\
2025-08-27T20:39:08 & -------- & 3$\times$180 s & -145.93 & 2.60 & 2.62 & -23.10 & 93.00 & 8.58 & 15.60 & 22.31 \\

2025-09-02T19:51:01 & 1500 s & 3$\times$180 s & -139.94 & 2.52 & 2.59 & -23.36 & 88.70 & 8.33 & 17.26 & 22.74 \\
2025-09-15T19:32:26 & 1500 s & 4$\times$180 s & -126.95 & 2.34 & 2.52 & -23.97 & 80.34 & 7.12 & 20.36 & 23.52 \\
2025-09-19T19:22:51 & 1500 s & 4$\times$180 s & -122.96 & 2.29 & 2.50 & -24.16 & 78.00 & 6.64 & 21.15 & 23.72 \\
\hline
\end{tabular}
\end{table*} 

\section{Gas Analysis} \label{sect:gas_analysis}

\begin{table}
\caption{Emission bands, and adopted $g$-factors and scale lengths at 1~au (in $10^{-13}$~erg~s$^{-1}$~mol$^{-1}$ and $10^{4}$~km, respectively). The $g$-factors are provided by \href{https://asteroid.lowell.edu/}{Lowell Minor Planet Services}(https://asteroid.lowell.edu/), with CN including its dependence on radial velocity. For CN, values span the range covered in the dataset. \resp{Scale lengths are adopted from \citet{1995Icar..118..223A}}}
\label{tab:gas_parameters}
\centering
\begin{tabular}{l c c c c}
\hline
Species & Spectral Region & $g$-factor & $l_p$ & $l_d$ \\
\hline
CN & 3830--3910 & 4.04 -- 4.36 & 1.3 & 21 \\
C3 & 3975--4150 & 10 & 0.3 & 2.7 \\
C2 ($\Delta \nu = 0$) & 4860--5210 & 4.5 & 2.2 & 6.6 \\
\hline
\end{tabular}
\end{table}

Our primary goal is to track the evolution of the gas production rates of the comet along the observing campaign and to constrain the onset of activity for the different molecular species in the visible. \resp{It is worth noting that we refer to the `onset' of activity as the earliest epoch in our dataset at which a given molecule becomes detectable above the sensitivity limits of our observations, rather than the physical onset (its presence in the coma).} On the night of September 19, observations using the B200 grism yielded no significant detection of the forbidden [O I] emission line that could be attributed to the comet, likely due to limited signal, contamination, or insufficient sensitivity.  As expected, the first molecular emission detected in our dataset is CN, with no detections of CN, C$_2$, or C$_3$ at heliocentric distances beyond 3.48~au, allowing us to place upper limits on their production rates. Given the low signal-to-noise level at the epochs prior to the onset of clear molecular emission, particular care was taken to ensure that the measured feature corresponds to genuine molecular emission rather than noise fluctuations. To this end, we adopt the following criteria: 

A detection is claimed only when (i) the integrated flux exceeds the $3\sigma$ noise level estimated from each molecular band, (ii) the spectral feature is consistent with the expected molecular band profile, and (iii) the emission is spatially centred on the comet optocenter. The $3\sigma$ threshold is derived by generating synthetic line profiles based on asymmetric Gaussian fits to cometary emissions observed with the same instrumental configuration \citep{2025MNRAS.538.1329M}. The peak of these profiles is then scaled to the $3\sigma$ noise level of each individual night, and the corresponding integrated flux is computed. A detection is considered to satisfy the $3\sigma$ criterion when the measured integrated flux within the molecular band exceeds that of the simulated profile. Nights when these criteria are not fulfilled are treated as non-detections and are used to place upper limits on the production rates. The same procedure is applied to all molecular species in the dataset (CN, C$_2$ and C$_3$), ensuring a homogeneous definition of detections and upper limits throughout the analysis.  While other species like NH$_2$ or CH remained below the $3\sigma$ noise level of the spectra, we did not estimate their upper limits because we lacked the empirical Gaussian profiles required for this calculation.

For both detections and upper limits, we converted integrated fluxes into total number of molecules ($N$) and production rates ($Q$) using a Haser model following e.g., \cite{2012ApJ...744....9H}, adopting the parent and daughter scale lengths listed in Table~\ref{tab:gas_parameters} and scaling them with $r_{\rm{h}}^2$. The reported uncertainties are formal and include only the propagated error from the measured CCD flux, assuming a linear relation between flux and derived quantities.

When the data quality allowed the extraction of resolved spatial profiles, production rates were derived from direct Haser model fits. The profiles were symmetric with respect to the nucleus, and both sides of the slit were combined to increase the signal-to-noise ratio. The analysis followed the methodology described by \cite{2025MNRAS.538.1329M}, including the same binning strategy, Monte Carlo simulations, and uncertainty estimation, and adopting the same parent and daughter scale lengths and $r_{\rm{h}}^2$ scaling. In this case, the uncertainty per bin was estimated as the maximum between the median absolute deviation and 20\% of the bin value. This percentage is derived from the relative dispersion measured in reliable bins, each containing 5–40 discrete data points (i.e., $N$-values sampled at a specific projected distance from the nucleus). For each night, the median dispersion within the bin was calculated. We then adopted the minimum among these median values as the final uncertainty estimates across the dataset.

\begin{figure}
    \centering
    \includegraphics[width=0.5\textwidth]{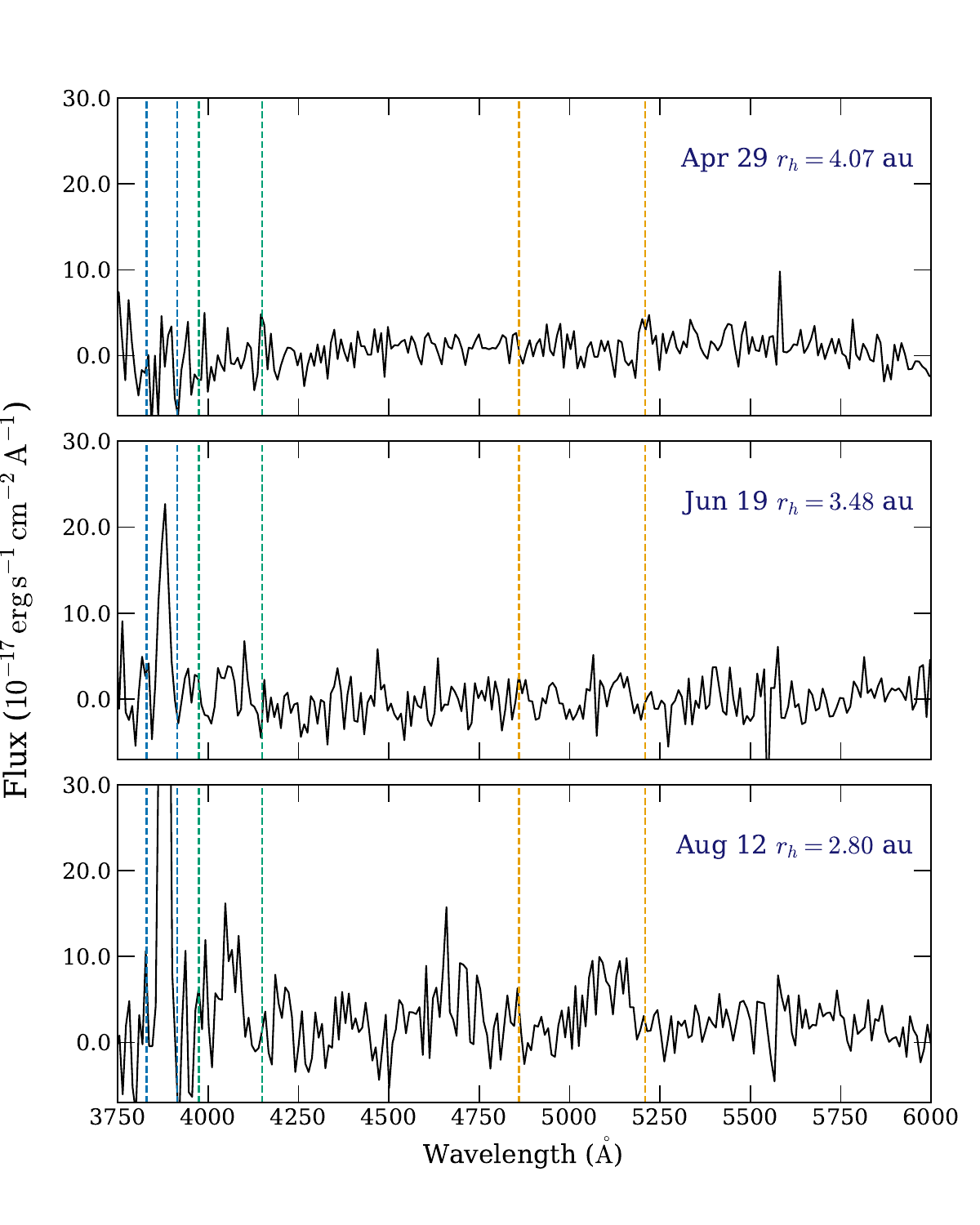}
    \caption{\resp{Flux-calibrated and dust-subtracted 1D spectrum obtained integrating the gas emission within a spatial window of $\pm50\,000$~km for three dates corresponding to no detection (top), first CN detection (middle), and first C$_2$ and C$_3$ detections (bottom). Vertical lines indicate the wavelength range used to measure the CN (blue), C$_3$ (green), and C$_2$ (yellow) emission bands.}} 
    \label{fig:cn_onset}
\end{figure}
\subsection{Gas production rates}

\begin{figure} 
   
    \centering
    \includegraphics[width=0.5\textwidth]{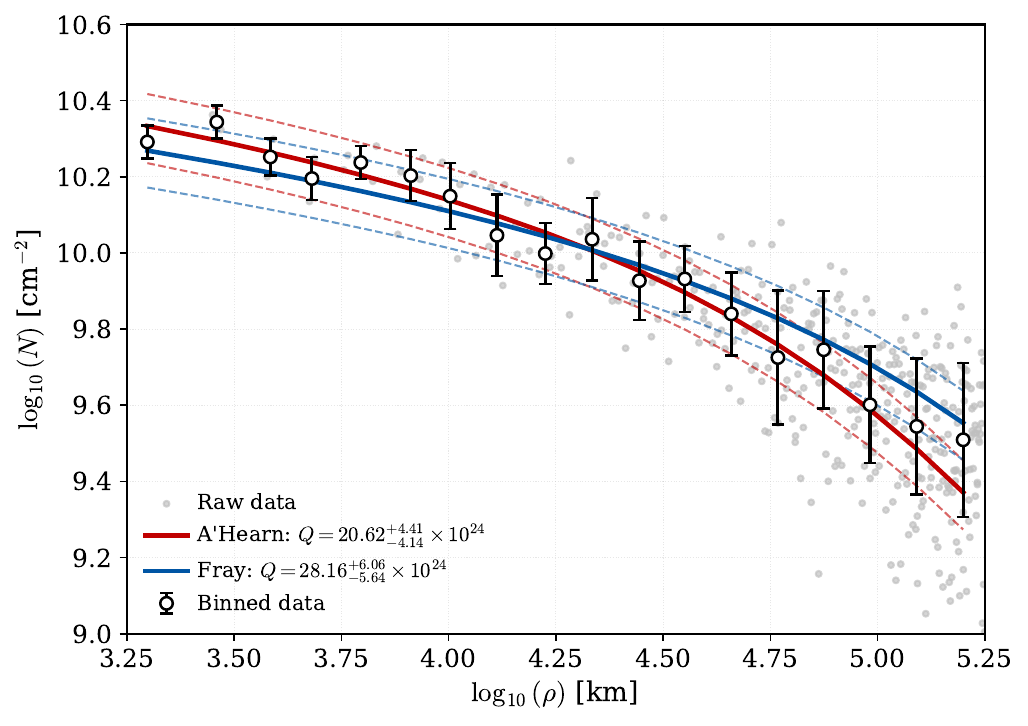}
    \caption{CN column density profiles for E1 on September 19 at $r_{\rm{h}} = 2.29$~au. Grey points represent the raw column density derived from spectra, while black open circles show the binned data and their uncertainties. The solid red line represents the best-fit Haser model using the empirical scale lengths from \protect\cite{1995Icar..118..223A}. The solid blue line shows the fit obtained using the scale lengths reported by \protect\cite{2005P&SS...53.1243F}, assuming HCN as the parent molecule. In both cases, dashed lines indicate the corresponding upper and lower $1\sigma$ uncertainties.}
    \label{fig:CN_profiles_months.pdf} 
\end{figure}

\begin{figure}  
    \centering
    \includegraphics[width=0.5\textwidth]{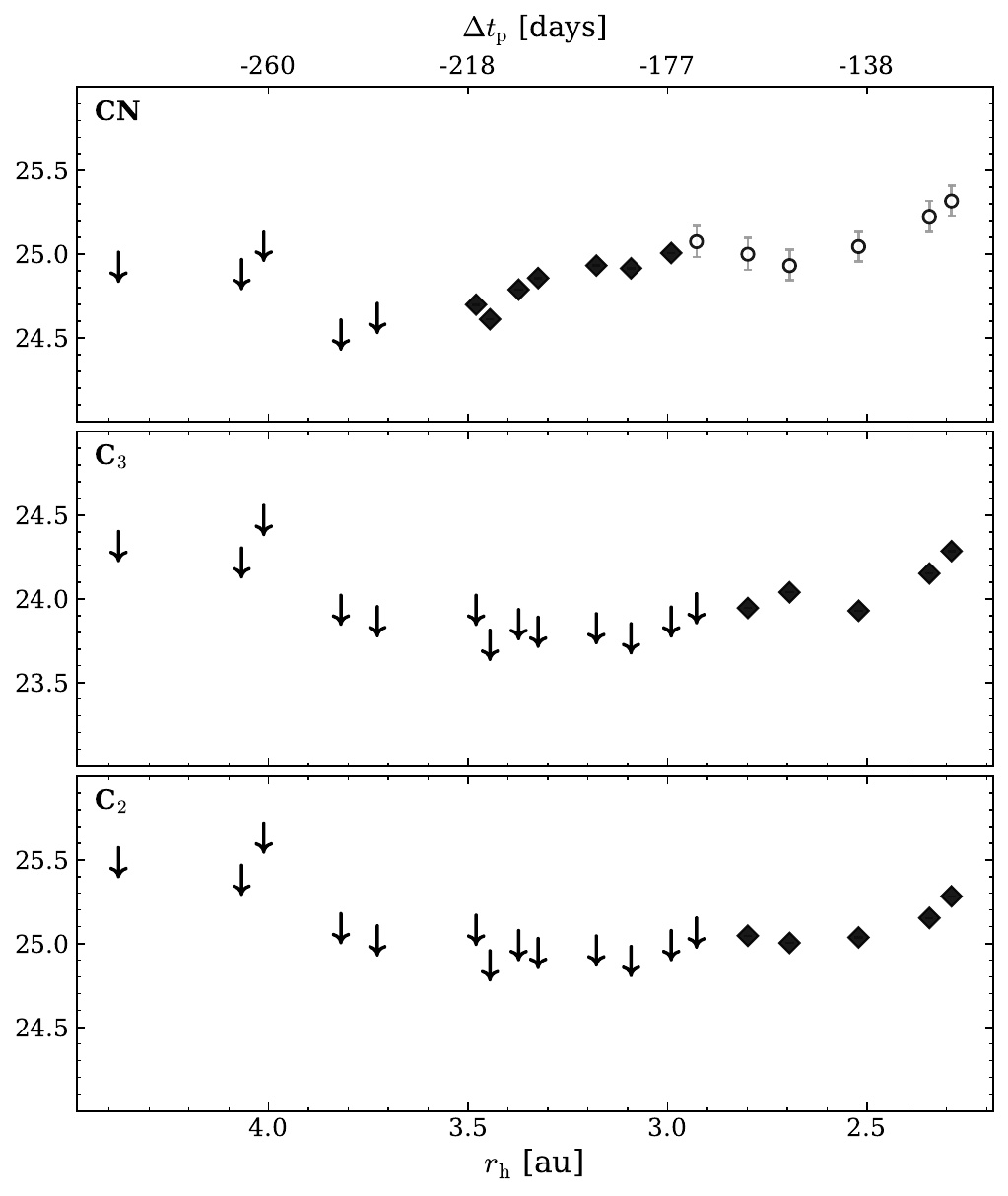} 
    \caption{Evolution of the gas production rates of E1 as a function of heliocentric distance \resp{($r_{\rm{h}}$)} for CN, C$_3$, and C$_2$ species (from top to bottom). \resp{The top horizontal axis displays the corresponding time to perihelion ($\Delta t_{\rm{p}}$) in days.} Diamonds correspond to integrated measurements, while open circles (shown only for CN at $r_{\rm{h}} < 3.2$ au) represent values derived from profile fitting. Vertical error bars are derived through standard error propagation. Upper limits are indicated with downward arrows.}
    \label{fig:gas_evolution_all_AHearn.pdf}
\end{figure}

\begin{table} 
\renewcommand{\arraystretch}{1.4}
\caption{Production rates derived in this work for CN, C$_3$, and C$_2$ in units of $10^{24}$~mol~s$^{-1}$. Upper limits are denoted with `<`. Values with errors are obtained by fitting the gas profile with the Haser model, while those without errors are obtained by integrating the signal within a rectangular aperture with a projected distance of 100,000~km for CN and 30,000~km for C$_2$ and C$_3$. In the latter case, formal uncertainties are estimated to be between 1 and 3$\%$.}
\label{tab:prod_rates}
\centering
\begin{tabular}{lccc}
\hline
Date & CN & C$_3$ & C$_2$ \\
\hline
2025-04-01 & $<8.40$ & $<2.07$ & $<30.66$ \\ 
2025-04-29 & $<7.59$ & $<1.65$ & $<24.05$ \\
2025-05-04 & $<11.22$ & $<2.96$ & $<42.99$ \\
2025-05-21 & $<3.31$ & $<0.86$ & $<12.36$ \\
2025-05-29 & $<4.16$ & $<0.74$ & $<10.46$ \\
2025-06-19 & $4.99$ & $<0.86$ & $<12.10$ \\
2025-06-21 & $4.09$  & $<0.53$ & $<7.42$ \\
2025-06-27 & $6.14$  & $<0.71$ & $<9.79$ \\
2025-07-01 & $7.19$  & $<0.64$ & $<8.78$ \\
2025-07-13 & $8.54$  & $<0.67$ & $<9.10$ \\
2025-07-20 & $8.23$  & $<0.58$ & $<7.87$ \\
2025-07-28 & $10.14$ & $<0.73$ & $<9.79$ \\
2025-08-02 & $11.88^{+2.7}_{-2.5}$ & $<0.88$ & $<11.66$ \\
2025-08-12 & $9.99^{+2.3}_{-2.1}$  & $0.88$ & $11.15$ \\
2025-08-20 & $8.55^{+1.83}_{-1.73}$ & $1.10$ & $10.10$ \\
2025-09-02 & $11.1^{+2.5}_{-2.2}$ & $0.85$ & $10.90$ \\
2025-09-15 & $16.8^{+3.6}_{-3.3}$ & $1.42$ & $14.24$ \\
2025-09-19 & $20.6^{+4.4}_{-4.1}$ & $1.93$ & $19.17$ \\
\hline
\end{tabular}
\end{table}
Following the criteria defined above, 
\resp{the first detection of} CN in our dataset occurs on June 19 when the comet was at $r_{\rm{h}} \sim 3.48$~au. Figure~\ref{fig:cn_onset} illustrates a sequence of E1 spectra before and after the onset. From this date until August 12, CN remained the only detectable molecular species. Between June 19 and August 2 production rates were derived by integrating the flux within a rectangular aperture corresponding to 100\,000~km centred on the nucleus of the comet. From August 2 onward, the CN emission spatially extended enough to allow the extraction of column density profiles, and production rates were obtained by fitting the observed emission with a Haser model, using the full spatial extent of the profile down to the noise level. Figure~\ref{fig:CN_profiles_months.pdf} shows an example of the CN column density and the best-fit model on September 19, when the comet was at $r_{\rm{h}} = 2.29$~au pre-perihelion. The fits were performed using the classical scale lengths from \cite{1995Icar..118..223A}, which we kept fixed for a reliable comparison with other studies. In fact, those standard scale lengths provide a good representation of the observed spatial profiles, better than shorter ones as described below.   

On the night of August 12, C$_2$ and C$_3$ emission were first detected in our dataset. For these species, as well as for the early CN observations prior to the availability of well defined spatial profiles, production rates were derived by integrating the flux within a rectangular aperture centred on the nucleus, adopting a projected distance of 100\,000~km for CN and 30\,000~km for C$_2$ and C$_3$. For all molecules, a constant gas outflow velocity of $v = 1$~km~s$^{-1}$
was adopted to derive production rates allowing a direct and consistent comparison with previous works. Figure~\ref{fig:gas_evolution_all_AHearn.pdf} shows the evolution of the different gaseous species as a function of heliocentric distance, while the corresponding production rates and upper limits are summarised in Table~\ref{tab:prod_rates}. 

CN production rates increase progressively toward the Sun, with a 
change in the $r_{\rm{h}}$ dependence around 2.5~au. A similar trend is observed for C$_2$ and C$_3$ from 2.5~au onward, though their behavior at larger distances remains unconstrained due to limited data. This transition, best captured by the extensive CN dataset, occurs shortly after water presumably begins to dominate gas production and may be related to a shift in the primary drivers of cometary activity.

We investigate the heliocentric dependence of the production rates assuming a power-law relation $Q \propto r_{\rm{h}}^{\,\alpha}$. As previously noted, the extensive CN dataset reveals a transition in activity between 2.7 and 2.5~au, with production rates being consistent with a flat behavior at $r_{\rm{h}} > 2.7$~au within uncertainties. Inward of 2.5~au, all three species show a synchronized and steep increase toward the Sun. While a fit restricted to this narrow range is statistically unreliable due to the limited number of data points (yielding extreme slopes of $-6.28$, $-5.32$, and $-8.13$ for CN, C$_2$, and C$_3$, respectively), it clearly indicates a 
change in the activity trend for all molecules.

To provide a more robust characterization, we re-derived the slopes using all detections for C$_2$ and C$_3$ ($r_{\rm{h}} < 2.8$~au) and, for consistency, recalculated the CN slope using this same subset. This yields more moderate values of $-3.98$ for CN, $-2.56$ for C$_2$, and $-3.21$ for C$_3$, which are in better agreement with the average slope of $\sim -2.75$ reported by \cite{1995Icar..118..223A} derived using a survey of comets at $r_{\rm{h}} < 3$~au. This transition likely marks a shift toward a regime dominated by water ice sublimation, showing that a single power law cannot capture the evolution of the comet as the primary drivers of activity change.

\section{Dust Modelling} \label{sect:dust}

\begin{figure}
    \centering
    \includegraphics[width=1\linewidth]{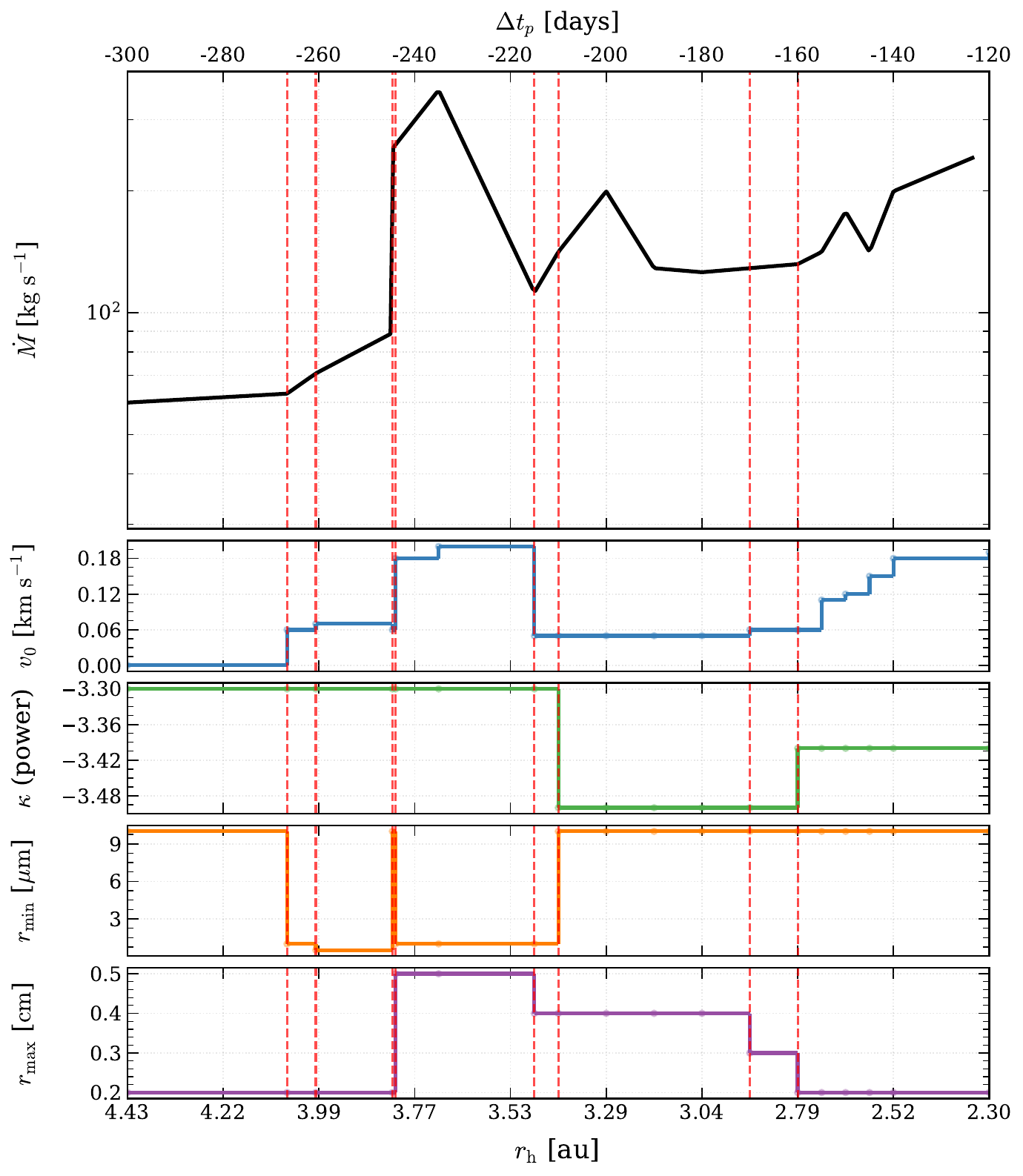}
    \caption{\resp{Dust environment parameters derived from the Monte Carlo modelling, shown as a function of heliocentric distance, $r_{\mathrm{h}}$ (for $r_{\mathrm{h}} \leq 4.43~\mathrm{au}$), and its respective time to perihelion, $\Delta t_{\mathrm{p}}$ ($\geq -300$~days, top axis). }
Top panel: time-dependent dust mass-loss rate, $\dot{M(t)}$ (kg\,s$^{-1}$, logarithmic scale). 
From top to bottom, the remaining panels display the reference ejection velocity parameter $v_0$, the particle size distribution index $\kappa$, and the minimum and maximum grain radii, $r_{\mathrm{min}}$ and $r_{\mathrm{max}}$ (m), adopted to reproduce the observed isophotes at each epoch. 
Vertical dashed lines mark the epochs at which changes were introduced in the particle size distribution parameters (i.e., in $\kappa$, $r_{\mathrm{min}}$, or $r_{\mathrm{max}}$).}
    \label{fig:model_params.pdf}
\end{figure}

\begin{figure*}
    \centering
    \includegraphics[width=1\linewidth]{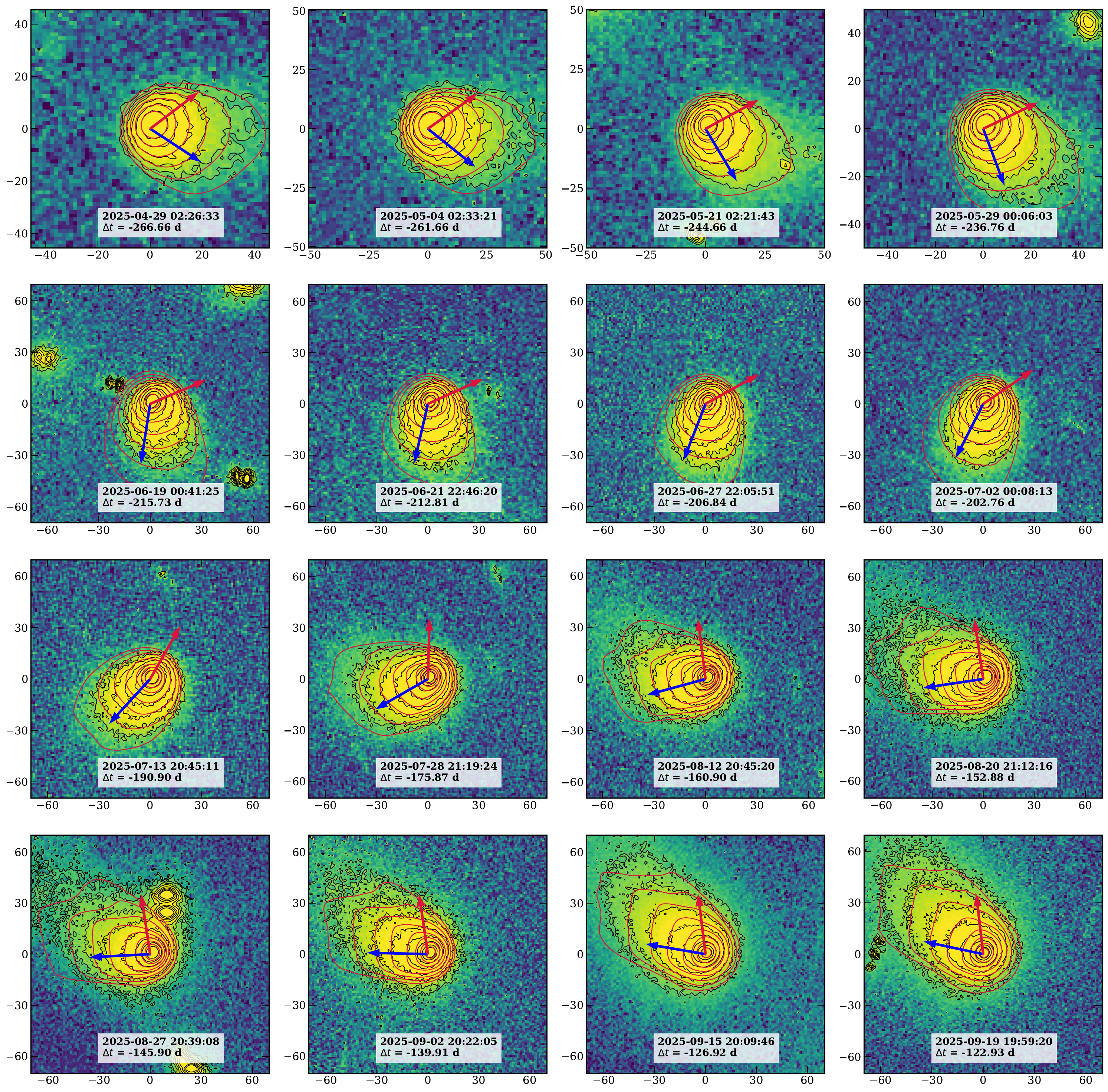}
    \caption{R-band images from our dataset with overlaid isophotes and best-fit results from the forward Monte Carlo dust model. The background shows the original R-band images, with their isophotes as black lines. Red lines are the corresponding isophotes obtained with the model. Both black and red contours correspond to the same level of surface brightness and are spaced by 0.5 magnitudes (with a range depending on the image). The X and Y axes indicate the projected distance from the nucleus in $10^3$ km. For each night, the date of observation (mean time of the images) and the time to perihelion in days are indicated. Blue and red arrows indicate the antisolar direction and the negative of the heliocentric velocity vector, respectively. North is up and East is to the left.}
    \label{fig:montecarlo.pdf}
    
\end{figure*}

\begin{figure}
    \centering
    \includegraphics[width=1\linewidth]{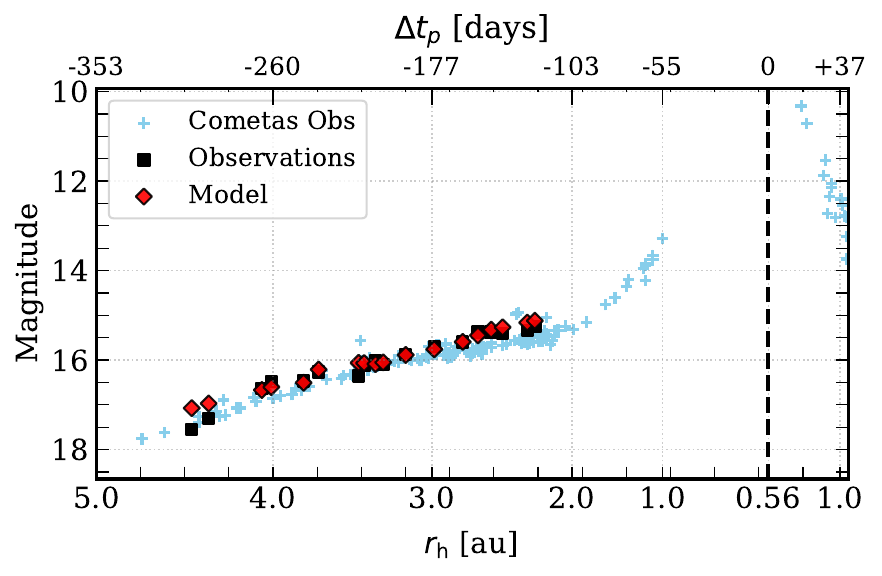}
    \caption{
    Evolution of the integrated magnitude using a $10'' \times 10''$ square aperture \resp{as a function of heliocentric distance, $r_{\mathrm{h}}$ and its respective time to perihelion, $\Delta t_{\mathrm{p}}$}. Vertical dashed line indicates the perihelion. Black squares denote our observational dataset, red diamonds indicate the Monte Carlo model results and blue cross symbols correspond with external estimates from \textit{Cometas\_Obs (http://www.astrosurf.com/cometas-obs/)}.}
    \label{fig:Mag_comparison_single}
\end{figure}

To characterize the dust environment of E1, we employed the Monte Carlo dust tail model by \cite{2025A&A...695A.263M}, publicly available on GitHub\footnote{\url{https://github.com/FernandoMorenoDanvila/COMTAILS/}}. This model has been extensively applied to retrieve the physical properties and mass-loss rates of ejected dust in various targets, including the well-studied comet 67P/Churyumov–Gerasimenko \citep{2017MNRAS.469S.186M}, the interstellar object 2I/Borisov \citep{2020MNRAS.495.2053D} and the distant comet C/2014 N3 (NEOWISE) \citep{2025A&A...697A.188I}, among others. In this work, we outline the most relevant aspects of the model, while a more detailed description can be found in \citet{2025A&A...695A.263M}. At a particular observational moment, the code calculates the position of a large ensemble of dust particles released from the nucleus earlier in time, and computes their combined brightness on the sky, enabling a direct, quantitative comparison with the observed images. This approach is particularly well-suited for the outer coma and tail, at distances greater than roughly 20 nucleus radii ($R_{\rm N}$), where the influence of nucleus gravity, particle collisions, and gas drag is negligible compared with solar gravity and radiation pressure.

The observed R-band signal is assumed to arise predominantly from sunlight scattered by dust particles, while contributions from gas emission are considered negligible. The dust environment is described by a large number of physical parameters. 
In particular, the time-dependent dust mass-loss rate, $\dot{M}(t)$, the particle size distribution, and the ejection velocity law. The size distribution of the ejected dust particles, defined by the radius $r$ assuming spherical grains, follows a power-law:
\begin{equation}
n(r)\,\propto \int_{r_{\mathrm{min}}}^{r_{\mathrm{max}}} r^{\kappa}\,\mathrm{d}r,
\end{equation}  
with minimum and maximum size limits $r_{\rm min}$ and $r_{\rm max}$. The motion of each particle is determined by the combined effects of solar gravity and radiation pressure. The ratio of these forces is expressed by the $\beta$ parameter: 
\begin{equation}
\beta = \frac{Q_{\rm pr}C_{\rm pr}}{2 \rho r},
\end{equation}  
 where $Q_{\rm pr}$ is the dimensionless scattering efficiency for radiation pressure, typically assumed to be $Q_{\rm pr} = 1$ \citep{1979Icar...40....1B}. $C_{\rm pr} = 1.19 \times 10^{-3}~\mathrm{kg~m^{-2}}$ \citep[e.g.,][] {1968ApJ...154..327F} is the radiation pressure coefficient, and $\rho$ is the particle density. 
 Smaller particles experience larger $\beta$ values and are therefore accelerated more strongly away from the Sun. The initial ejection velocity is assumed to depend on particle size, heliocentric distance $r_{\rm{h}}$, and the local solar illumination conditions. It is parameterized as

\begin{equation} \label{eq:velocity_general}
v_{\rm ej}(r, r_{\rm{h}}, z) = v_0 \, \beta^{\gamma} \, r_{\rm{h}}^{-\delta} \, \cos^{x} z,
\end{equation}

where $v_0$ is a reference velocity, $\gamma$ governs the dependence on particle size through the radiation pressure parameter $\beta$, and $\delta$ describes the scaling with heliocentric distance. The term $\cos^{x} z$ accounts for the dependence of the ejection velocity on the solar zenith angle $z$, with $x$ controlling the strength of this angular modulation. This formulation captures the expected trends of smaller particles being ejected faster and of comet activity declining with increasing heliocentric distance.

Assuming a spherical nucleus, the particles can be ejected in three different regimes: isotropic, preferentially sunward, or concentrated in a localized active area. As stated above, each particle is propagated from its release time to the epoch of observation, and its position is projected onto the plane of the sky. Its brightness contribution is calculated from its scattering cross-section and the observing geometry. The contributions from all simulated particles are then accumulated to construct a synthetic image of the coma and tail, which is directly compared with the observed images. Instrumental effects, including pixel scale and point-spread function, are incorporated to match the observational setup.

Given the large number of free parameters inherent to the Monte Carlo dust model, the solution cannot be considered strictly unique. We therefore present a solution that consistently reproduces the observed brightness levels and morphological evolution across the full dataset. Certain parameters are fixed based on prior studies. 
\resp{Thus, in line with \citet{2025MNRAS.541L...8S}, we set the geometric albedo of the nucleus to 0.04 and its phase function coefficient to 0.047 \citep[see, e.g.,][]{2024come.book..361K}. For the ejected particles, we also assume a geometric albedo of 0.04 and adopt the phase function provided by \href{https://asteroid.lowell.edu/comet/dustphase.html}{Schleicher and Marcus (2010)}\footnote{\url{https://asteroid.lowell.edu/comet/dustphase.html}}.} Other magnitudes were fixed to physically reasonable values, specifically a nucleus radius of 1~km and a particle bulk density of $\rho = 1000$~kg~m$^{-3}$. The remaining parameters were constrained by comparison with the data. 

As a first step, we determined the magnitudes that govern the ejection velocity, finding the best-fit dependencies on particle size and heliocentric distance to be $\gamma = 0.35$ and $\delta = 0.5$, respectively, with a solar zenith angle modulation exponent of $x = 0.4$. These values were held constant throughout the campaign, and the expression for the ejection velocity (Eq.~\ref{eq:velocity_general}) for E1 becomes:

\begin{equation}
v_{\rm ej}(r, r_{\rm{h}}, z) = v_0 \, \beta^{0.35} \, r_{\rm{h}}^{-0.5} \, \cos^{0.4} z. 
\end{equation}

Once this velocity expression was established, the other magnitudes were refined through a global iterative fit to match the observations at each epoch. In this process, the dust production rate $\dot{M}(t)$ was treated as a fully free, time-dependent parameter, starting from an initial asymmetric Gaussian approximation to estimate the global activity trend. To achieve the best agreement with the observed isophotal structure and tail morphology, we also allowed temporal adjustments in the reference velocity $v_0$ (between 0.05 and 0.2~km~s$^{-1}$), the particle size distribution index $\kappa$ (from $-3.3$ to $-3.5$), and the grain radii limits, $r_{\rm min}$ (0.5--10~$\mu$m) and $r_{\rm max}$ (0.2--0.5~cm).

Using the aforementioned derived parameters, we generated synthetic images of E1 throughout the observing period. Each simulation incorporated 100 directional events per particle, 1000 temporal bins, and 100 spatial bins, resulting in a total of $10^7$ particles propagated from the onset of activity to the epoch of each observation. 
For the simulations, the dust ejection was assumed to occur preferentially in the sunward direction, and the onset of activity was fixed at 2024 February 15, corresponding to the date of the earliest available observation of the comet. The synthetic images were then compared with the observed images to assess the agreement in coma brightness and tail morphology. The evolution of the time-dependent parameters is shown in Figure~\ref{fig:model_params.pdf}, while the corresponding synthetic images are compared with the observed isophotes in Figure~\ref{fig:montecarlo.pdf}.

The retrieved dust production curve (Figure~\ref{fig:model_params.pdf}, top panel) shows a peak of enhanced activity around \resp{May 22 ($r_{\rm{h}}\sim 3.8$~au, $\Delta t_{\rm{p}} \sim -244$~days)}, possibly indicating a short-lived increase in dust emission. However, we cannot confirm a corresponding rise in gas production due to the absence of measurements for this period. Our photometric data for these dates show no evidence of a sharp brightness increase; the observed magnitude follows a roughly linear trend, consistent with the coma magnitude estimated from the Monte Carlo model. The model reproduces the observations well around \resp{May 22}. The closest observational epoch after this increase is May 29, which shows an excellent agreement between the modelled and observed magnitudes, both yielding a value of 16.55 within a projected aperture of 10,000~km, as also illustrated in Figure~\ref{fig:montecarlo.pdf}. Examination of the lower panels in the same figure shows that, during this period, the reference ejection velocity $v_0$ increased, and the minimum and maximum grain radii, $r_{\rm min}$ and $r_{\rm max}$, were larger than at other epochs. These changes indicate that faster and larger particles were emitted, which exit the CCD-covered region more quickly and contribute less per particle to the integrated brightness. Consequently, despite the enhanced dust production suggested by the model, the observed magnitude remains relatively constant. Nevertheless, we cannot rule out that this feature may partly reflect edge effects in the modelling, particularly given that this epoch lies at the very beginning of our observational coverage. Consequently, the results derived from modelling the earliest observations (\resp{specifically those taken where $r_{\rm{h}} \gtrsim 3.5$~au, corresponding to $\Delta t_{\rm{p}} < -220$~days}) should be interpreted with caution, as they are more susceptible to these boundary artifacts. \resp{We performed alternative runs of the code and found that reasonable isophotal profiles can be obtained even without the initial peak by introducing small adjustments to the power-law index $\kappa$ and the minimum and maximum grain radii. Crucially, these alternative configurations lead to variations of less than 10\% in the dust production rate within our most reliable region ($\Delta t_{\rm{p}} \ge -220$~days). However, we still prefer our original solution (with the apparent `outburst') as it provides the overall best fit to the observations across the entire dataset.}

In the most reliable region of our model, both dust production and ejection velocity exhibit a moderate and steady increase as the comet approaches perihelion. The dust production rate peaks at approximately 240~kg~s$^{-1}$; a value remarkably low for a DNC. For instance, this mass-loss rate is significantly lower than the one reported for C/2022~E3 (ZTF) at similar heliocentric distances \citep{2024A&A...683A..51L}, even though that comet was already considered to have low activity compared to typical DNCs. This low dust production is consistent with the weak gas emission and the overall low level of activity observed for E1 throughout our campaign (see Sect.~\ref{sect:discussion}). Simultaneously, the terminal ejection velocity factor reaches a peak of $\sim$180~m~s$^{-1}$, resulting in maximum emission velocities of approximately 20 and 100~m~s$^{-1}$ for large and small particles, respectively, values reported at the subsolar point. When accounting for the $\cos(z)^{0.4}$ dependence and increasing $r_{\rm{h}}$ values, the actual emission velocities are significantly lower.

Regarding grain properties, our solution yields a power-law size distribution with a slope of $-3.5$, slightly flattening to $-3.4$ as the comet approaches perihelion. We also find that the minimum grain radius remains stable at $\sim$1~$\mu$m, while the maximum radius ($r_{\rm max}$) shows a notable decreasing trend from 5~mm to 1~mm toward perihelion. This behaviour may be explained by intensified surface erosion as solar heating increases; as sublimation fronts move closer to the surface, reduced cohesion or rapid fragmentation of the topmost layers could favour the emission of smaller average particle sizes.

For a further cross-check of our results, Figure~\ref{fig:Mag_comparison_single} presents a comparison between our integrated magnitudes, the Monte Carlo model predictions, and independent reports from the \textit{Cometas\_Obs} Association\footnote{\url{http://www.astrosurf.com/cometas-obs/}}. For consistency, all magnitudes were derived using a $10'' \times 10''$ square aperture to match the standard used by independent observers. A remarkable agreement is found between our observations and the model results, both of which follow the trend reported by external observers.

\resp{A key challenge in models with a large number of magnitudes is how the assumed physical properties of the comet impact our derived dust production rates. We performed sensitivity tests by varying, in particular, the nucleus size and the density of the ejected particles ($\rho$). The results show that the chosen nucleus size has a negligible effect on the extended coma simulations, as it only impacts the pixels immediately surrounding the optocenter. Conversely, variations in $\rho$ directly affect the simulated isophotes. In particular, when keeping our baseline best-fit parameters unchanged and varying $\rho$ between $500$ and $2000~\text{kg m}^{-3}$, we obtain a similar fit to the observations by shifting the absolute dust production rate globally by factors of $0.7 \times dM/dt$ and $1.6 \times dM/dt$, respectively. Alternatively, this density effect can be formally compensated by varying the velocity scaling factor ($v_0$, see Figure~\ref{fig:model_params.pdf}). Changes in $\rho$ shift the terminal velocities, according to equation~3,  $v_{\rm{ej, new}}/v_{\rm{ej, ref}} = (\rho_{\rm{ref}}/\rho_{\rm{new}})^{0.35}$. Formally, if we adjust the velocity scaling factors to compensate for this density-driven change ($v_0 \times 1.27$ for $\rho = 500~\text{kg m}^{-3}$ and $v_0 \times 0.78$ for $\rho = 2000~\text{kg m}^{-3}$), the trajectories of the particles remain unaltered. This yields identical isophotes without requiring any modification to the dust production rate. We constrain the derived dust production rate within $0.7 \times dM/dt$ and $1.6 \times dM/dt$ when accounting for variations in the assumed particle density.}

\section{Discussion} \label{sect:discussion}

\subsection{Gas Activity}

The discovery of E1 at 8.2~au initially pointed to a highly active object. However, our monitoring reveals that this comet has a moderate-to-low activity. Comparing for example with C/2020~V2, \cite{2025MNRAS.543.1178A} obtained that, within the range of common heliocentric distances, this comet had productions rates one order of magnitude larger than those of E1.   Thus E1 awakens early but fails to develop a high-productivity coma, leaving species like NH$_2$ or CH below the $3\sigma$ noise level of the spectra. Despite this, it is noteworthy that E1 displayed detectable CN production as far as 3.48~au, indicating that early CN release is not exclusive to "hyper-active" comets like C/1995 O1 (Hale-Bopp), but a more common feature among DNCs. Such deviations from early expectations are not unique. For instance, comet C/2025 L1 similarly exhibited a rapid and unexpected decline in brightness following its perihelion passage. These cases underscore the inherent unpredictability of these pristine targets and the critical need for systematic monitoring of DNCs. Ultimately, understanding why targets like E1 deviate from expected activity is fundamental to characterizing the diversity of the Oort Cloud and optimizing the scientific return of future flyby missions such us \textit{Comet Interceptor} mission (ESA), which rely on long-term observational baselines to distinguish truly promising targets from those with inhibited activity. \resp{Moreover, given the nature of the \textit{Comet Interceptor} mission, 
the final target will be selected with limited prior observational data. 
Therefore, characterizing similar objects that show early activity becomes more 
important, as these observations can be used as reference baselines to interpret 
the data that will be available during the initial mission phases.}

 \resp{We compare our results with other comets that exhibit activity in this regime. For instance, early CN emissions at similar heliocentric distances have been documented in other DNCs, such as C/2017 K2 and C/2020 V2, with the latter showing a CN production rate of $\sim 6.9 \times 10^{25}$~mol~s$^{-1}$ \citep{2025PSJ.....6..272C, 2025MNRAS.543.1178A}. Among periodic objects, \cite{2024MNRAS.534.1816F} reported a production rate of $2.2 \times 10^{25}$~mol~s$^{-1}$ for comet 12P/Pons-Brooks at $r_{\rm{h}} = 3.5$~au. However, classical surveys using similar techniques (ground-based long-slit spectroscopy and imaging) show a lack of observations at large heliocentric distances \citep[e.g.,][]{1995Icar..118..223A, 2009Icar..201..311F, 2011Icar..213..280L, 2012Icar..218..144C}. For instance, only two of the 26 comets in \citep{2011Icar..213..280L} were observed at $r_{\rm{h}} > 3$~au, with only one (Hale-Bopp) showing CN emissions. Similarly, \cite{2009Icar..201..311F} presented a survey of 92 comets, where less than 10\% of the total observations were taken at $r_{\rm{h}} > 3$~au. even of these objects showed CN emissions at $r_{\rm{h}} > 3.4~au$, two periodic comets (D/1993 F2-A at 5.36~au and 29P/Schwassmann-Wachmann 1 at 5.77~au) and five long-period comets / DNCs at $r_{\rm{h}}$ between 3.4 and 3.9, which exhibited CN activity levels between $10^{25}$ and $10^{26}$~mol~s$^{-1}$. Our detection of CN in E1 at 3.48 au directly adds to this growing collection of distant data, and may indicate that production of CN beyond 3.5~au is likely a common feature that has probably been underreported due to a lack of dedicated observational campaigns. Furthermore, our derived CN production rate at 3.48~au ($\sim 5 \times 10^{24}$~mol~s$^{-1}$) is relatively low compared to the aforementioned comets, but aligns with the overall low activity and the low dust production derived in Section~\ref{sect:dust}.}

To further characterize the gas environment, we recalculated $Q_{\text{CN}}$ assuming HCN as the primary parent molecule. For this purpose, we adopted the HCN scale lengths ($l_p$) and the $r_{\rm h}$-dependent velocity expression from \cite{2005P&SS...53.1243F}. This allows us to retrieve absolute $Q$ values rather than comparative $Q/v$ ratios.

A comparison between both sets of scale lengths shows they reproduce the observed profiles with comparable quality. However, as $r_{\rm{h}}$ decreases, the empirical scales from \cite{1995Icar..118..223A} provide a marginally better fit than those from \cite{2005P&SS...53.1243F}. This is illustrated in Figure~\ref{fig:CN_profiles_months.pdf}, which displays the profile at the lowest $r_{\rm{h}}$ ($2.29$~au); even here, where the discrepancy between models is most noticeable, both fits remain remarkably consistent with the data. This is particularly evident considering the inherent noise at larger projected distances ($\log(\rho) > 4.5$).

Our results suggest that at larger heliocentric distances ($\sim 3$~au), the data are fully compatible with CN originating from HCN sublimating directly from the nucleus. However, the trend toward shorter scale lengths as the comet approaches the Sun suggests a more complex environment. This may indicate the emergence of additional parent species or a contribution from distributed sources, such as CN produced directly from dust grains. These results are consistent with \cite{2005P&SS...53.1243F}.

\subsection{Taxonomy of E1}

\begin{figure}  
    \centering
    \includegraphics[width=\linewidth]{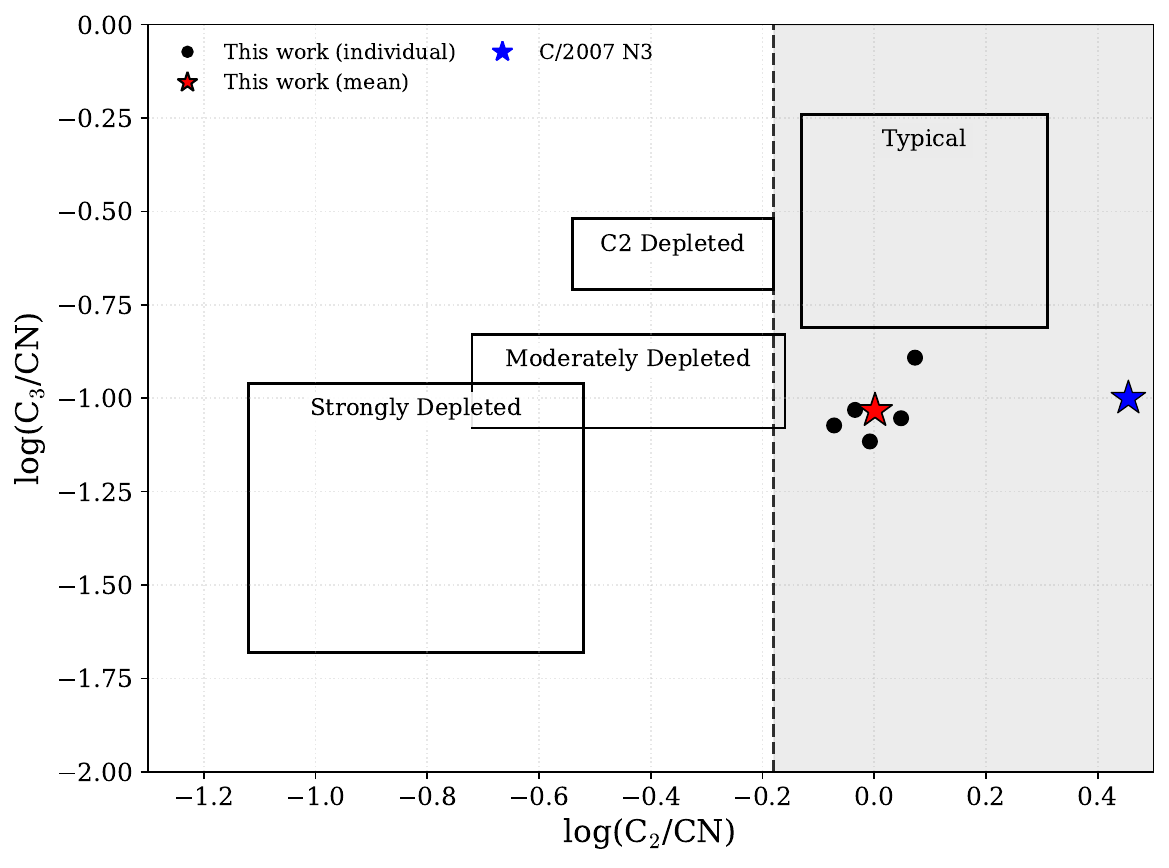}

     \caption{Logarithmic abundance ratios of C$_2$ and C$_3$ relative to CN. The black points represent individual measurements of E1 from this work, while the red star indicates the corresponding sample mean. The blue star corresponds to measurements of comet C/2007 N3 reported by \protect\cite{2011AJ....141...12B}. The bounded regions indicate the taxonomic classification scheme based on carbon-chain depletion, including strongly depleted, moderately depleted, C$_2$-depleted, and typical comets, as defined by \protect\cite{2025PSJ.....6..248B}. The dashed vertical line marks $\log(\mathrm{C}_2/\mathrm{CN}) = -0.18$, the threshold for carbon-chain depletion established by \protect\cite{1995Icar..118..223A}, the grey dashed region indicating the typical compositional range.}
     \label{fig:test_taxonomic}
\end{figure} 

The taxonomic classification of a comet is based on molecular abundance ratios and is therefore sensitive to the adopted molecular scale lengths, which can affect the derived production rates and, consequently, the resulting ratios. For this reason, consistent scale lengths are essential when comparing with previous studies. In this work, we adopt the same set of scale lengths as in \cite{1995Icar..118..223A} and \cite{2025PSJ.....6..248B}, which allows for a direct and meaningful comparison. In particular, the taxonomy defined by \cite{1995Icar..118..223A}, based on a survey of 85 comets (41 with well determined production rates of several gas species) classifies objects into typical and carbon-chain depleted comets. This framework has been extended by \cite{2025PSJ.....6..248B}, who, using a larger sample of 220 comets (135 in what they call the restricted subset of well determined production rates of five gas species), introduce additional subclasses within the depleted population, such as strongly depleted, moderately depleted, and C$_2$-depleted comets. We compute the molecular abundance ratios $\log(\mathrm{C}_2/\mathrm{CN})$ and $\log(\mathrm{C}_3/\mathrm{CN})$, hereafter referred to as C$_2$/CN and C$_3$/CN, respectively. Using this approach, we obtain C$_2$/CN values ranging from $-0.072$ to $0.073$, with a mean value of $0.00 \pm 0.07$, and C$_3$/CN values ranging from $-1.16$ to $-0.89$, with a mean value of $-1.03 \pm 0.09$. The quoted uncertainties correspond to the standard deviation of the measurements. All the derived molecular abundance ratios are shown in Figure~\ref{fig:test_taxonomic}, where the taxonomic classification regions from aforementioned surveys are also indicated. 

These values are in good agreement with the typical range defined by \cite{1995Icar..118..223A}, where comets have $\mathrm{C}_2/\mathrm{CN} \approx 0.06 \pm 0.10$ (and carbon-chain depleted comets fall below $-0.18$). Our measured C$_2$/CN value, including its dispersion, encompasses the typical value, supporting the classification of E1 as a typical comet within their scheme (grey dashed area shown in Figure~\ref{fig:test_taxonomic}). These authors reported that the ratios of production rates, Q(C$_2$) and Q(C$_3$) relative to Q(CN), typically vary by less than a factor of 2 for individual comets, with a slightly larger dispersion observed for Q(C$_3$)/Q(CN). In our dataset, the maximum variation between individual measurements corresponds to factors of less than 1.5 for Q(C$_2$)/Q(CN) and about 1.7 for Q(C$_3$)/Q(CN), for observations taken on different nights spanning heliocentric distances over $\sim 0.5$~au.

Comparing our measurements with the taxonomic regions defined by \cite{2025PSJ.....6..248B} (bounded boxes in Figure~\ref{fig:test_taxonomic}), E1 is placed in a mixed regime. Specifically, the C$_2$/CN ratio is consistent with the typical comet population, while the C$_3$/CN value falls within the moderately depleted region, lying below the expected range for typical compositions. This behaviour suggests a relative depletion in C$_3$ with respect to CN when compared to the general trends observed in large cometary samples. Similar patterns have been reported in the literature. For instance, \cite{2011AJ....141...12B} (blue star in Figure~\ref{fig:test_taxonomic}) using the same scale lengths, found C$_2$/CN to be 0.43 and C$_3$/CN values of -1 for comet C/2007 N3, indicating a composition relatively enhanced in C$_2$ and depleted in C$_3$, placing that object also outside the different taxonomic classes defined by the authors. This comparison indicates that, although unusual, the compositional behaviour observed in E1 is not unique and may reflect a broader continuity in carbon-chain abundances among comets, with gradual transitions between the traditionally defined taxonomic categories. Within this framework, all derived ratios for E1 fall within the region between typical and moderately depleted comets, and we find no evidence for any internal evolutionary or transitional behaviour within a single object. Instead, E1 appears to occupy an intermediate compositional regime, with C$_3$ being slightly depleted while C$_2$ remains typical.

To compare with other surveys and different choices of scale lengths, we recompute all molecular abundance ratios using the set of scale lengths from \cite{2012Icar..218..144C}, following the same procedure as in our nominal analysis. According to these authors, comets are classified as carbon-chain depleted when C$_2$/CN $< 0.02$ and C$_3$/CN $< -0.86$. We obtain mean values of $0.10 \pm 0.07$ for C$_2$/CN and $-1.10 \pm 0.08$ for C$_3$/CN. Applying these scale lengths, we recover the same C$_3$-depleted and C$_2$-typical pattern derived using the framework of \cite{2025PSJ.....6..248B}. This identifies E1 as a "mixed" taxonomic object, a chemical signature that remains robust regardless of the specific modelling parameters or classification criteria employed. Notably, while the C$_2$/CN ratio is classified as typical, it lies very close to the depletion threshold, suggesting that the composition of E1 is placed on the boundary between these taxonomic groups.

We also derive the molecular ratios using the scale lengths from \cite{2011Icar..213..280L}, scaled following their reported power-law dependence. However, these results should be interpreted with caution, as the CN spatial profiles in our dataset are not well reproduced when using these scale lengths and their variation with $r_{\mathcal{h}}$, which directly impacts the reliability of the derived production rates and, consequently, the molecular ratios. Despite this limitation, we compare our results within the taxonomy framework where C$_2$/CN and C$_3$/CN are used to classify comets according to their dynamical group.

In this taxonomy scheme, dynamically new comets typically show values of C$_2$/CN $\sim 0.2 \pm 0.2$ and C$_3$/CN $\sim -0.9 \pm 0.1$. For E1, we obtain mean values of $-0.27 \pm 0.05$ for C$_2$/CN and $-1.42 \pm 0.07$ for C$_3$/CN, indicating that the abundances of this comet do not fit within the dynamically new comet class under this classification. In particular, the C$_3$/CN ratio is significantly lower than the mean values reported for all dynamical classes (Jupiter-family, Halley-type, long-period, and dynamically new comets), whereas the C$_2$/CN ratio is closer to the values typically found for long-period comets ($\sim -0.1 \pm 0.1$). To place this result in context, \cite{2021MNRAS.507.5376I} reported mean values of $-0.15 \pm 0.10$ for C$_2$/CN and $-1.86 \pm 0.10$ for C$_3$/CN in the dynamically new comet C/2019 Y4, obtained using the same set of scale lengths. Their results also show a markedly low C$_3$/CN ratio, even lower than the value derived here, while the C$_2$/CN ratio remains comparatively higher, supporting the idea that such behaviour is not unique.

\subsection{Af$\rho$ as proxy of dust production}

We calculated the $Af\rho$ parameter as a function of projected aperture $\rho$ using
\begin{equation}
Af\rho = \frac{4 r_{\rm{h}}^2 \Delta^2}{\rho} \, 10^{-0.4 (m_{\rm c} - m_\odot)},
\end{equation}
where $r_{\rm{h}}$ and $\Delta$ are the heliocentric and geocentric distances, $m_{\rm c}$ is the observed magnitude of the comet, and $m_\odot$ is the apparent solar magnitude in the same band. The observed profiles indicate that $Af\rho$ remains approximately constant for $\rho \gtrsim 10{,}000$~km, suggesting that the surface brightness of the coma decreases as $1/\rho$ at these distances from the nucleus, a trend which our Monte Carlo model reproduces. Consequently, we adopt the $Af\rho$ value at this distance as representative of the coma's brightness scaling. To obtain the zero-phase values, $A(0)f\rho$, we applied a phase-angle correction using the phase function defined by \href{https://asteroid.lowell.edu/comet/dustphase.html}{Schleicher and Marcus (2010)}\footnote{\url{https://asteroid.lowell.edu/comet/dustphase.html}}, as implemented in the \texttt{sbpy} Python package \citep{2019JOSS....4.1426M}. $A(0)f\rho$ is traditionally employed in the literature as a standard proxy for cometary dust production. Our estimates of $A(0)f\rho$ remain remarkably flat throughout the observing period, averaging approximately 810~cm. When comparing this value with other comets observed at similar heliocentric distances, we find a high degree of variability. For instance, according to \cite{2022ATel15822....1J}, comets C/2020 V2 (ZTF) and C/2021 Y1 (ATLAS) exhibited $A(0)f\rho$ values of $9089\, \pm\, 1553$~cm and $470\, \pm\, 15$~cm at $r_{\mathrm{h}} = 2.76$ and $2.56$~au, respectively. This highlights that the $Af\rho$ is highly comet-dependent.

\resp{We evaluate the behavior of the $Af\rho$ proxy alongside our Monte Carlo results to examine how each approach tracks the comet's evolution. While the $Af\rho$ proxy has the advantage of involving fewer free parameters than our numerical model, it exhibits a slight decrease as the comet approaches perihelion. In contrast, the model-derived dust production rate shows a steady increase over the same period (see Figure~\ref{fig:model_params.pdf}, top panel). Even if the initial peak around May 22 is treated as a potential artifact, the rising trend in $dM/dt$ persists.}

To contextualize our results, we compared our dataset with independent observations from the \textit{Cometas\_Obs} Association. For this purpose, we recalculated our $Af\rho$ values to match their specific nightly apertures, accounting for the fact that the exact physical distance ($\rho$) varies depending on the comet-observer distance. In this comparison, our derived $Af\rho$ values are systematically higher than those reported by the association; the origin of this discrepancy remains unclear. Conversely, our results show good agreement with \cite{Bryssinck2026}, who reported an apparent magnitude of 15.89 (Bessel-R filter) within a $10.54''$ aperture and an $Af\rho$ of 496~cm at $\rho = 10{,}000$~km. For the same physical scale, our closest observing epochs (May 4 and 21, 2025) yielded $Af\rho$ values of 504~cm and 469~cm, respectively. The minor discrepancy of approximately 0.15~mag is likely due to inherent differences in photometric calibration.

\subsection{Dust-to-gas mass ratio} 
\begin{figure}
    \centering 
    \includegraphics[width=0.48\textwidth]{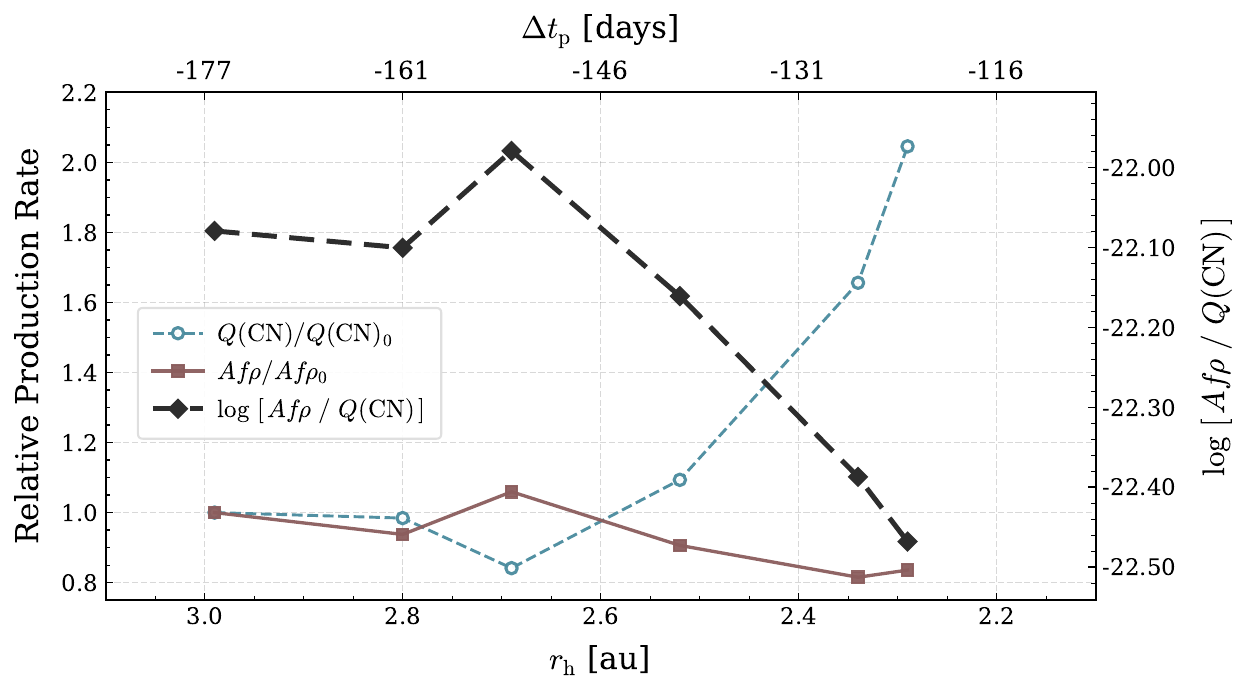}
    \vspace{0.3cm}  
    \includegraphics[width=0.48\textwidth]{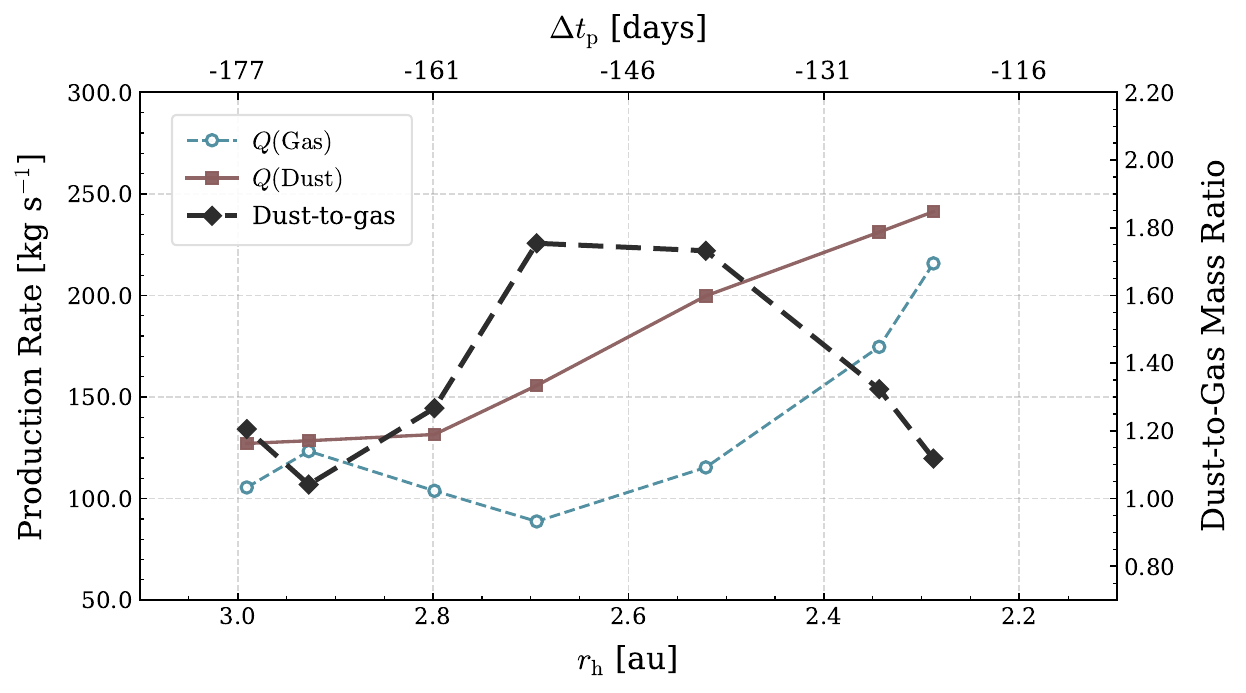}
   \caption{\resp{Evolution of the dust-to-gas ratio and individual production components as a function of heliocentric distance $r_{\mathrm{h}}$ and time relative to perihelion $\Delta t_{\mathrm{p}}$ (top axes). 
\textit{Top panel:} Relative production rates (left axis) of the CN production rate (open blue circles, dashed line) and the $A(0)f\rho$ dust parameter (filled brown squares, solid line), both normalized to their respective values observed at $\sim 3\mathrm{~au}$. The right axis displays the logarithmic ratio $\log [A(0)f\rho / Q(\mathrm{CN})]$ (black diamonds, long-dashed line). 
\textit{Bottom panel:} Absolute production rates in $\mathrm{kg\ s^{-1}}$ (left axis) for the gas ($\rm{Q}_{\mathrm{H}_2\mathrm{O}}$, open blue circles, dashed line) and dust derived from the Monte Carlo dust tail modelling ($\dot{M}_{\mathrm{dust}}$, filled brown squares, solid line) components. The right axis shows the dust-to-gas mass ratio (black diamonds, long-dashed line).}}
    \label{fig:dust_to_gas_comparison}
\end{figure}

The dust-to-gas ratio is frequently given as the $\log[Af\rho/Q(\mathrm{CN})]$. For the range of heliocentric distances between 3 and 2.29~au, our smallest distance, that ratio varies between -21.98 and -22.47. Considering the values reported in \cite{1995Icar..118..223A} from their ensemble of comets, E1 would be in the upper area of their data, suggesting that E1 would be a dust-rich comet. Specifically, at $r_{\mathrm{h}} \sim 2.80$~au, E1 shows a ratio of approximately $-22.10$, which is slightly higher than the $-22.26$ reported for comet C/2020 V2 (ZTF) at a similar distance of 2.76~au. Similarly, at $r_{\mathrm{h}} \sim 2.52$~au, E1 maintains a ratio of $-22.16$, again exceeding the value of $-22.52$ observed for C/2021 Y1 (ATLAS) at 2.56~au \citep{2022ATel15822....1J}. In both cases, the higher $\log[Af\rho/Q(\mathrm{CN})]$ values of E1 compared to these targets indicate a greater relative dust content, suggesting that E1 is a dust-rich comet within the considered population.

Notably, we observe a clear decreasing trend in the $\log[Af\rho/Q(\mathrm{CN})]$ ratio as the comet approaches the Sun. This behavior is a direct consequence of the different evolutionary scales of the dust and gas production: while our estimation of $A(0)f\rho$ estimates remain nearly flat throughout the period, CN production rate shows a steady increase with decreasing $r_{\mathrm{h}}$. Consequently, the relative proportion of dust to gas appears to drop, although this likely reflects the limitations of $Af\rho$ as a mass-loss proxy rather than a physical depletion of the dust reservoir.

As we lack direct measurements of either OH or CO$_2$ production throughout the campaign, dust to gas ratio can only be roughly estimated. Thus, to  derive the total gas mass-loss, we have to assume that the gas production is dominated by water where $r_{\rm{h}} < 3$~au, using the CN production rates as a proxy via an adopted CN--OH conversion. For this purpose, we use the relation of \cite{1995Icar..118..223A}, appropriate for typical comets, which provides $\log[Q(\mathrm{OH})/Q(\mathrm{CN})] \approx 2.5$. Within this framework, E1 is classified as a typical comet, and this conversion is taken as our baseline. The water production rate is then obtained from OH using a standard photodissociation branching ratio, $Q(\mathrm{H_2O}) \approx 1.1\,Q(\mathrm{OH})$. Dust production rates are derived independently from our Monte Carlo dust model. Both gas and dust production rates are converted into kg\,s$^{-1}$ to enable a direct comparison. It is worth noting that since the CN--OH conversion is empirically established for heliocentric distances $r_{\rm{h}} < 3$~au, we restrict the dust-to-gas analysis to this "safe" range. This ensures that the water-driven activity baseline is applied consistently where it is most reliable, avoiding the uncertainties of the transitional regime sampled at larger distances. 

\resp{Figure \ref{fig:dust_to_gas_comparison} shows the evolution of both dust-to-gas ratios: the observational proxy ($\log[A(0)f\rho/Q(\text{CN})]$, top panel) and the dust-to-gas mass ratio ($M_{\text{dust}}/M_{\text{gas}}$, lower panel). We take the water production inferred from $Q$(CN) as the gas mass loss rate ($M_{\text{gas}}$). Inside 2.7~au, both ratios exhibit a localized decreasing tendency as the comet approaches perihelion, driven by the fact that the gas activity rises much faster than the dust production. However, this behavior is suggested only from three data points (three nights of observations), so we cannot establish any clear trend.} We obtain results for the dust-to-gas mass ratio of E1 ranging between 1 and 1.8 over the observing campaign. No clear dependence on heliocentric distance is observed within the sampled range. This suggests that, within the uncertainties of the analysis, both dust and gas production evolve in a similar manner over time, without a dominant trend in their relative contribution as a function of heliocentric distance.

It is important to note that the dust-to-gas ratios reported here should be regarded as estimates, as they rely on an indirect determination of the water production rate from CN and therefore depend on both the adopted CN–OH conversion and the assumed scale lengths. In particular, different calibration choices can lead to non-negligible variations in the absolute values. For instance, using the scale lengths and the CN--OH conversion factor of 2.17 derived by \cite{2012Icar..218..144C} for long-period comets (instead of the customary value of 2.5) yields dust-to-gas ratios in the range of 1.8 to 3.0 for the same dataset. This discrepancy illustrates the high sensitivity of the absolute dust-to-gas values to the initial assumptions and highlights the importance of using a consistent framework when comparing different objects. In both cases, the derived ratios remain consistently above or near unity, indicating that the mass loss in E1 is dominated by the dust component rather than by the volatile species. This is in agreement with the A(0)f$\rho$ results, which also point toward a dust-dominated coma. However, we find a key difference between the two approaches: while the $\log[Af\rho/Q(\mathrm{CN})]$ ratio shows a decreasing trend as the comet approaches perihelion, our model-derived dust-to-gas mass ratio does not exhibit such a decline, probably due to an artifact of the $A(0)f\rho$ proxy.

\section{Summary and conclusions}  

In this work, we present a comprehensive pre-perihelion characterization of E1, employing the Haser model to analyze the gas production rates from multi-epoch long-slit spectroscopy, and a Monte Carlo simulator to characterize the dust activity and mass-loss rates from R-band imaging. Our observations, spanning heliocentric distances from 4.5 to 2.3~au, reveal that the object exhibits activity well before the classical water-sublimation regime. \resp{We detected CN activity at 3.48 au, a distance where such emissions have been previously observed in other comets across different families. Our results add to the population of comets observed beyond 3 au, which represent less than 10\% of the total entries in classic spectroscopic surveys.} 
This \resp{detection} 
could be explained by the presence of a volatile-rich surface layer that remains relatively unprocessed. The analysis of the CN spatial profiles shows that the gas distribution is physically consistent with HCN acting as the primary parent molecule, especially at $r_{\rm h} \sim 3$~au, whereas at lower $r_{\rm h}$, more compact scale lengths provide a marginally better fit. This trend suggests the emergence of a more complex coma environment, potentially involving additional parent species or contributions from distributed sources, such as dust grains. 

The taxonomic classification based on molecular production ratios places the comet in an intermediate compositional regime. While the C$_{2}$/CN ratio is consistent with the typical population, the C$_{3}$/CN ratio indicates a moderate depletion, placing the object on the boundary between the typical and moderately depleted classes. This mixed signature suggests that the chemical diversity of Oort Cloud comets presents a wide range of properties that may overlap traditional boundaries. Such behavior suggests the existence of a compositional continuum, where individual objects like E1 exhibit intermediate characteristics that bridge the gap between established taxonomic groups. 

The Monte Carlo dust tail model successfully reproduces the observed coma morphology, with a peak mass-loss rate of approximately 240~kg~s$^{-1}$. According to our simulations, the dust environment is characterized by a power-law size distribution with a slope of $\sim -3.5$ and a constant minimum grain radius of $r_{\text{min}} = 1$~$\mu$m, values that can be considered typical among comets \citep[see, e.g.][]{2024come.book..653A}. Interestingly, the maximum grain size, $r_{\text{max}}$, decreases from 5~mm to 1~mm as the comet approaches the Sun, a trend that may point toward enhanced surface erosion or the fragmentation of larger aggregates under increasing solar insolation. Furthermore, the terminal ejection velocities, peaking at $\sim$100~m~s$^{-1}$ for the smallest grains at the subsolar point, follows a $v \propto \beta^{0.35} r_{\rm{h}}^{-0.5} \cos^{0.4}(z)$ dependence.

We also want to emphasize that the traditional $Af\rho$ parameter is inherently limited as a mass-loss proxy, as it relies on numerous oversimplifications regarding grain size distributions and ejection velocities. Consequently, it does not provide a realistic representation of the actual dust production rate, which is better captured by detailed dynamical modeling. Within the range where water is assumed to be the dominant volatile ($r_{\rm{h}} < 3$~au), we derive a dust-to-gas mass ratio that consistently remains above unity, ranging from $\sim$1 to 3 depending on the choice of scale lengths and CN--OH conversion. This result indicates that the mass loss in E1 is slightly dominated by the refractory component, characterizing it as a ``dust-rich'' object. Ultimately, these findings provide a critical framework for understanding the diverse activity regimes of Oort Cloud comets and offer valuable insights for the \resp{planning} and future interpretation from the \textit{Comet Interceptor} mission.

\section*{Acknowledgments}
Based on observations collected at Centro Astronómico Hispano en Andalucía (CAHA) at Calar Alto, proposals 25A-2.2-018 and 25B-2.2-008, operated jointly by Junta de Andalucía and Consejo Superior de Investigaciones Científicas (IAA-CSIC). We thank the amateur association \texttt{Cometas\_Obs} for providing us with aperture photometry data for comet C/2024 E1 (Wierzchos), particularly to Felipe G\'omez Pinilla.

The authors acknowledge financial support from grants PID2021-126365NB-C21, PID2024-156684OB-I00, and from the Severo Ochoa grant CEX2021-001131-S funded by MICIU/AEI/10.13039/501100011033. IME acknowledges financial support from the FPI grant PRE2022-105422 funded by MICIU/AEI/10.13039/501100011033 and by European Social Fund Plus (ESF+). 

The Monte Carlo dust tail code used to generate the synthetic images makes use of the JPL-Horizons online ephemeris system. 
\section*{Data Availability}
The data underlying this article will be shared on reasonable request to the corresponding author.



\bibliographystyle{mnras}
\bibliography{example} 




\appendix

 

\bsp	
\label{lastpage}
\end{document}